\documentclass[iop]{emulateapj}
\usepackage{epstopdf}

\shortauthors{Ruan, McQuinn, \& Anderson}
\shorttitle{Detection of Quasar Feedback}
\begin{document}
\title{Detection of Quasar Feedback from \\ the Thermal Sunyaev-Zel'dovich Effect in \emph{Planck}}
\author{John~J.~Ruan\altaffilmark{1,2}, 
  Matthew~McQuinn\altaffilmark{2}, 
  Scott~F.~Anderson\altaffilmark{2}
  }
\altaffiltext{1}{ Corresponding author: jruan@astro.washington.edu} 
\altaffiltext{2}{Department of Astronomy, University of
Washington, Box 351580, Seattle, WA 98195, USA}

\begin{abstract}
	Poorly understood feedback processes associated with highly-luminous black hole accretion in 
quasars may dramatically affect the properties of their host galaxies. We search for the effect of 
quasar feedback on surrounding gas using \emph{Planck} maps of the thermal Sunyaev-Zel'dovich effect (tSZ). 
By stacking tSZ Compton-y maps centered on the locations of 26,686 spectroscopic quasars from the Sloan 
Digital Sky Survey, we detect a strong but unresolved tSZ Compton-y signal at $>$5$\sigma$ significance
that likely originates from a combination of virialized halo atmosphere gas and quasar feedback effects. 
We show that the feedback contribution to our detected quasar tSZ signal is likely to dominate over 
virialized halo gas by isolating the feedback tSZ component for high- and low-redshift quasars. We find that 
this quasar tSZ signal also scales with black hole mass and bolometric luminosity, all consistent with general 
expectations of quasar feedback. We estimate the mean angularly-integrated Compton-y of quasars 
at $z\simeq1.5$ to be 3.5$\times$10$^{-6}$ Mpc$^2$, corresponding to mean total thermal energies in 
feedback and virialized halo gas of $1.1\,(\pm 0.2) \times 10^{62}$ erg, and discuss the implications for quasar 
feedback. If confirmed, the large total thermal feedback energetics we estimate of 5\% ($\pm1\%$ statistical 
uncertainty) of the black hole mass will have important implications for the effects of quasar feedback on 
the host galaxy, as well as the surrounding intergalactic medium.
\end{abstract}

\section{Introduction}	
	The coevolutionary view of galaxies and their central supermassive black holes (SMBHs), suggested 
by the connection between SMBH masses and the velocity dispersion of their host galaxy 
bulges (the $M_{\rm{BH}}$-$\sigma$ relation) \citep[e.g.,][]{fe00, ge00, me01}, has highlighted 
the important role of black hole feedback in the evolution of massive galaxies 
\citep[for a recent review, see][]{fabian12}. Feedback from the quasar phenomenon is one particular 
form of active galactic nuclei (AGN) feedback, triggered by large-scale inflows of gas towards the central SMBH, 
possibly due to major galaxy mergers \citep{sa88, ba97, ho06}. The parsec-scale accretion process onto 
the SMBH is radiatively efficient and highly luminous, with bolometric luminosities of order 10$^{45-47}$ erg s$^{-1}$, 
close to the Eddington limit. Radiative line-driving \citep{mu95, pr00} or magnetic forces \citep{pr03, ev05, fu14} 
are believed to inevitably drive winds, commonly observed as blueshifted ($\Delta v \sim10,000$ km s$^{-1}$)
absorption lines in the ultraviolet \citep[e.g.][]{we91, gi09, bo13}, and in the X-rays with 
near-relativistic blueshifts \citep[e.g.][]{re09, char09, to12}. These winds can drive out interstellar gas 
from the the host galaxies, resulting in the galactic-scale ($\sim$10 kpc) quasar outflows now frequently observed 
\citep{al10, st11, greene11, ru11, gr12, ha12, ma12, li13, ha14}. This removal of gas from the host galaxies
can limit the gas inflow to their nuclei, thus self-regulating accretion onto their central SMBH
\citep[e.g.][]{si98, fabian99}. 

	A possible consequence of quasar feedback is the observed quenching of 
star formation in the host galaxy bulge \citep{farrah12, pa12, la14}. This coupling of black hole growth 
to star formation can naturally lead to observed correlations between properties of the host galaxy and the 
SMBH mass \citep{ki03}, and can also explain the similar observed redshift evolution of star formation and 
quasar activity in galaxies \citep[e.g.][]{sc04}. Many of these observed trends that may owe to quasar 
feedback have been reproduced in semi-analytic models and cosmological simulations of galaxy formation
that utilize various sub-grid models to approximate quasar feedback 
\citep[e.g.][]{ka00, di05, sp05, bo06, si07, so08}. Despite these advances, because quasar 
activity predominantly occurs at high redshifts and is often obscured by dust and gas, details of the feedback 
energetics, mechanism, and effects on galaxy evolution are still poorly understood.

	Several analytical studies and simulations have suggested that the effects of 
quasar feedback on the host galaxy interstellar medium (ISM) and surrounding circumgalactic medium 
can be probed using the thermal Sunyaev-Zel'dovich effect \citep[tSZ,][]{su70, su72} in the Cosmic Microwave 
Background (CMB) with the current generation of arcminute-scale angular resolution CMB experiments 
\citep{vo94, na99, pa02, la03, ch07, sc08, ch08}. Since the tSZ effect is a direct measurement of the total 
thermal energy in free electrons, a detection would provide a valuable probe of quasar feedback energetics. 
However, the quasar feedback tSZ signal predicted by these studies is too small to be 
detected for individual quasars. Instead, a cross-correlation (or stacking) approach between CMB maps and 
large samples of quasars is required with current CMB measurements to achieve high signal-to-noise 
detections. This stacking approach is taken here.

	 \citet{ch10} reported a tentative detection of quasar feedback through the tSZ effect by 
cross-correlating photometrically-selected SDSS quasar candidates with \emph{WMAP} CMB maps;
this detected quasar tSZ signal was at least an order of magnitude larger than that predicted from 
cosmological simulations tuned to match the $M_{\rm{BH}}$-$\sigma$ relation. However, the origin of 
the \citet{ch10} tSZ signal is ambiguous, as it may be due to ionized gas in the quasar host galaxies 
heated by quasar feedback and/or virialized halo atmospheres. Indeed, the hot intracluster medium 
(ICM) in massive galaxy clusters, originally heated by virial shocks during the cluster formation 
process, dominates the tSZ sky. Aside from the ICM of galaxy clusters, there is now evidence from the 
tSZ effect that virialized halo atmospheres also exist in the halos of more moderate-mass galaxies. 
By stacking the tSZ signal from galaxies binned in stellar mass $M_\star$ using \emph{Planck} CMB maps, 
\citet[][hereafter PC13a]{planck13a} report robust ($>$3$\sigma$) detection of the tSZ effect in galaxies 
with halo masses down to log $\,M_{\mathrm{halo}} = 13.2$ $M_{\odot}$ \citep[also see][]{gralla14}. 
Intriguingly, PC13a find that the amplitude of this tSZ signal scales as a power-law from clusters to more 
modest-sized galaxies (the $Y-M_\mathrm{halo}$ relation), extending from 
clusters down to galaxies that are nearly three orders of magnitude smaller in halo mass
\citep[although see][]{greco14}.

	Separating the quasar feedback and virialized halo gas contributions to the quasar tSZ 
signal can be difficult without knowledge of the exact quasar host halo mass distribution at the redshift of 
the quasar sample. Large-scale clustering measurement have established that quasars lie in halos with 
a range of masses with mean log $\,M_{\mathrm{halo}}\simeq 12.4-12.9$ $M_{\odot}$ \citep[e.g.][]{po04, 
cr05, my06, sh07, da08, ro09, wi12, fo13}. This canonical mean quasar halo mass is nearly independent of 
redshift over the redshift range we probe ($z \lesssim 3$), and has only a weak dependence on physical 
parameters such as the back hole mass, $M_{\rm{BH}}$, or the bolometric luminosity, $L_{\rm{bol}}$ 
\citep{sh09, tr12}. Although the tSZ signal from virialized gas in halos of $\sim$$10^{12.7}$ $M_\odot$ 
(the fiducial mean quasar halo mass) is expected to be negligible based on the power-law $Y-M_\mathrm{halo}$ relation 
reported by PC13a, a small (and poorly-constrained) number of quasars lying in larger, cluster-mass halos can 
significantly increase the stacked mean quasar tSZ signal, thus biasing the inferred quasar feedback energetics.

	Our approach to isolating the feedback tSZ signal is to measure the quasar tSZ 
signal for a subsample of high-redshift ($z\simeq2$ ) quasars. Since cluster-mass halos effectively do not exist 
at high redshifts, the mean tSZ signal for this subsample is not expected to be contaminated from the ICM of 
the small number of cluster-mass halos. To confirm our results, we will also compare our high-redshift quasar tSZ
signal to that from a separate subsample of low-redshift ($z\simeq1$) quasars. We will show that the low-redshift
quasar tSZ signal is roughly consistent with that expected from a combination of (1) quasar feedback with similar 
feedback parameters as for high-redshift quasars, and (2) a non-negligible tSZ signal from virial halo gas that is 
consistent with estimates based on the quasar host halo mass distribution.

	In this paper, we will utilize a tSZ Compton-y map produced using CMB maps from 
\emph{Planck}. Aside from the greatly improved angular resolution and sensitivity in comparison to 
\emph{WMAP}, \emph{Planck}'s large number of frequency channels (six in the High Frequency Instrument) 
is well-suited for tSZ studies. These channels span a wide frequency range, including a band centered 
approximately at the 217 GHz tSZ anisotropy null, as well as an 857 GHz band that is dominated by thermal 
dust emission. This enables the spectral separation of signal components, and thus high signal-to-noise 
detections of the tSZ effect. 
	
	The outline of this paper is as follows: In Section 2, we provide a brief primer on the tSZ effect and quasar 
blastwaves, and then we describe the \emph{Planck} tSZ Compton-y map, dust-correction procedure, and 
the quasar and galaxy samples we utilize. Section 3 describes the tSZ map stacking process, the resulting 
quasar and galaxy tSZ signals, and validation of our stacking procedure through comparisons to previous 
\emph{Planck} results. In Section 4, we discuss details of the observed quasar tSZ signal, the inferred total 
thermal energetics, dependence on quasar physical properties, and implications for quasar feedback. 
We summarize and conclude in Section 5. Throughout this paper, we assume a standard $\Lambda$CDM 
cosmology with $\Omega_\mathrm{m} = 0.272$, $\Omega_\Lambda = 0.728$, and $H_0 = 70.4$ km s$^{-1}$ 
Mpc$^{-1}$, consistent with the \emph{WMAP}7 results of \citet{ko11}.
	
\section{Analysis Methods and Data}

\subsection{The Thermal Sunyaev-Zel'dovich Effect}
	In the tSZ effect, inverse Compton scattering of CMB photons by thermal motions of energetic 
electrons induces a distortion on the blackbody CMB spectrum along the line of sight of the form 
\begin{equation}
	\frac{\Delta T}{T_0} = y~[x~\mathrm{coth}(x/2) -4],
\end{equation} 
	at dimensionless frequency $x = h\nu/k_\mathrm{B}T_0$ \citep{su70, su72}, where 	
$T_0$ is the temperature of the CMB \citep[for a review, see][]{ca02}. 
Equation 1 neglects the small relativistic corrections to this spectral distortion, which are only important at extremely high temperatures ($>10^8\;$K).
This spectral distortion is parameterized by the dimensionless tSZ Compton-y parameter integrated along 
any line of sight $l$ in direction $\mathbf{\hat n}$
\begin{equation}
	y(\mathbf{\hat n}) = \int n_e \sigma_T \frac{k_\mathrm{B} T_e}{m_e c^2}\,\mathrm{d}l,
\end{equation} 
where $n_e$ and $T_e$ are the number density and temperature of the scattering electrons, respectively.

	The total tSZ signal for a source at redshift $z$ is characterized by the angularly integrated Compton-y parameter
\begin{equation}
	Y(z) = D_{\mathrm{A}}^2(z) \int y(\mathbf{\hat n}) \, \mathrm{d} \Omega 
\end{equation} 
integrated over some solid angle $\Omega$ of sky and expressed in Mpc$^2$, where $D_{\mathrm{A}}(z)$ is the 
angular diameter distance. This integration of the Compton-y over solid angle captures the 
integrated Compton-y in a cylinder along the line of sight.
 
	After measuring $Y(z)$ as described by Equation 3, the implied total thermal energy in electrons within this radius is then
\begin{equation}
	E_{\mathrm{e}} = \frac{3}{2} Y(z) m_e c^2 / \sigma_T.
\end{equation} 
For protons and electrons in thermal equilibrium, the total energy in ionized gas is simply
\begin{equation}
	E_{\mathrm{tot}} = (1 + \frac{1}{\mu_e}) E_{\mathrm{e}},
\end{equation} 
where $\mu_e$ is the mean particle weight per electron, for which we will assume the primeval value of 1.17. 

	Due to the differences in angular diameter distance, our comparison of the mean integrated 
Compton-y's both internally between different redshift bins and to those of PC13a (to validate our stacking 
procedure) is aided by a rescaling of the tSZ signal to a common $D_{\mathrm{A}} = 500$ Mpc. We will aim to 
show that the detected quasar tSZ signal is inconsistent with simply originating from virialized halo gas, which 
is the dominant signal measured in PC13a. Thus, it is also helpful to divide out the expected redshift dependence 
of the integrated Compton-y of virialized gas in galaxies, which we can do if we assume self-similar redshift 
scaling of isothermal virialized halo gas in hydrostatic equilibrium. Following PC13a, we thus define the rescaled 
integrated Compton-y:
\begin{equation}
	\tilde{Y} =  \left( \frac{Y(z)}{D_{A}^2(z)} \right) E^{-2/3}(z) \left(\frac{D_{\mathrm{A}}(z)}{500~\mathrm{Mpc}} \right)^2,
\end{equation} 
expressed in arcmin$^2$, where $E(z)^2 = \Omega_\mathrm{m}(1+z)^3 + \Omega_\Lambda$. 
For self-similar scaling and if the virialized halo gas dominates the tSZ, the quantity $\tilde{Y}$ will be
constant with redshift at fixed halo mass.

	Although there are likely to be correlations between quasars and other CMB secondary anisotropies, they do not significantly affect our analysis or results. Specifically, in the stacking approach we take, the kinetic SZ effect \citep{su80} will average to zero due to the random line of sight velocities of quasars. Similarly, for gravitational lensing of the CMB \citep{bl87} by quasar host halos, the sign of the lensing anisotropy is dependent on whether hot or cold regions of the CMB are lensed, which average to zero in a stack. Finally, our estimate of the non-linear integrated Sachs-Wolfe effect \citep{rees68} shows that it is negligible for quasar host halos.

\subsection{Quasar Wind Bubbles}

	Quasars are known to drive winds that can be comparable energetically to their electromagnetic 
emissions. At the shocked interface between the wind bubble and the ambient medium, roughly half of the 
energy in the outflow is thermalized \citep{oskrikermkee} and can hence be visible in a Compton-y map. 
If the wind bubble is energy conserving, the physics of the bubble would depend little on the energy injection 
process and would extend into the circumgalactic medium (and eventually the intergalactic medium). The radius
of this blastwave would be \citep{koomckee92}
\begin{equation}
R_{\rm BW} \approx  0.2 \, E_{62}^{1/5} t_{8}^{2/5} Z_{1.5}^{-3/5} \Delta_{100}^{-1/5} ~{\rm Mpc},
\label{eqn:rS}
\end{equation}
where $\Delta_{100}$ is the mean enclosed pre-shock gas density in units of $100$ times the cosmological 
mean, and $Z_{1.5} = (1+z)/2.5$. Equation~(\ref{eqn:rS}) assumes a continuous energy injection over a 
time $t_{8}$ with luminosity $E_{62}/t_{8}$, where $E_{62}$ is the total injected energy in units of 
$10^{62}$~erg, and $t_{8}$ is the time over which the feedback energy is injected in units of $10^8$~yr. 
Quasar lifetime estimates favor $t_{8} = 0.1-10$ (e.g. \citealt{martini01}), and the reference energy of 
$10^{62}$~erg is $5\%$ of the rest mass energy of a $10^9M_\odot$ black hole, a number we will revisit later
\footnote{If we had instead assumed all of the energy had been injected instantaneously a time 
$t_{8}$ ago (which is the Sedov-Taylor blastwave), the shock radius $r_s$ would be larger by a factor 
of $2$.  This limit might apply if most feedback occurred during an earlier time, such as an obscured 
phase \citep{hopkins}.}. For the $10'$ FWHM resolution of the {\it Planck} Compton-y map used here 
(Section~2.3), a distance of $\{3.7, 4.8, 5.1\}$ Mpc is resolved for $z=\{0.5, 1, 1.5\}$, respectively, 
distances much greater than the expected $R_{\rm BW}$ while the quasar is active. Hence, the 
Compton-y profile around quasars is unlikely to be resolved in this study.

	Equation~(\ref{eqn:rS}) is valid if the injected energy is not radiated away and is hence energy 
conserving. \citet{fau12} argued that this would be the case for quasar winds. In particular, they showed that at 
small radii, where the inverse-Compton cooling time is short, the proton-electron equilibration time is 
long enough that the bulk of the energy is carried by the protons, making quasar winds energy conserving
(at least for fast winds with injection velocities $>$$10^4$ km s$^{-1}$). In order for the wind energy to later 
produce an observed Compton-y signal, the proton bath needs to recouple to the electrons well 
downstream where the cooling time is long. We find that this is likely as the electron-proton 
equilibration timescale via Coulomb scattering at $R_{\rm BW}$ is
\begin{equation}
t_{\rm ep} 
            \approx 5 \; E_{62}^{3/5} Z_{1.5}^{-9/5} t_8^{-9/5} \Delta_{100}^{-3/5}~~{\rm Myr},
            \label{eqn:tep}
\end{equation}
By evaluating $t_{\rm ep}$ at the blastwave temperature of $k_b T_{BW} = 3/32 \mu_e m_p \dot{R}_{BW}^2$ 
and at a post-shock density enhancement of four as implied by the Rainkine-Hugoniot jump conditions for a 
strong adiabatic shock. Equation 8 ignores conduction, which would act to further reduce $t_\mathrm{ep}$.
Thus, $t_{\rm ep}$ is smaller than the quasar lifetime $t_{8}$ for a range of plausible parameters, 
motivating searches for the imprint of quasar feedback through the tSZ effect around quasars.

	This picture of energy-driven winds is supported by recent investigations of the driving of 
galactic-scale outflows by quasar winds using idealized and cosmological simulations by \citet{co14}, who 
showed that the momentum-flux from energy-conserving winds are required to match observed outflows. 
Furthermore, X-ray observations of quasars also do not appear to show signatures of strong inverse-Compton 
cooling of the shocked wind \citep{bou13}, suggesting that vast amounts of thermal energy lies in the ionized gas surrounding quasars. 

\subsection{Planck Compton-y Map}

	We utilize the publicly available tSZ Compton-y map of 
\citet[hereafter HS14]{hi14}\footnote{http://www.astro.princeton.edu/~jch/ymapv1/}, produced using 
15.5-month nominal mission data from \emph{Planck}. We only briefly summarize its characteristics
relevant for our investigation and refer the reader to HS14 for more details on its construction and 
validation. This Compton-y map was created using a modified constrained Internal Linear Combination 
\citep[ILC,][]{re11} method, using the 100, 143, 217, 353, and 545 GHz channel CMB maps 
from \emph{Planck}'s High Frequency Instrument (HFI). In this modified ILC algorithm, the observed emission
in the channel maps are assumed to be linear combinations of blackbody primary CMB emission, tSZ spectral
distortion (parameterized by the tSZ Compton-y in Equation 1), and noise. Specifically, the CMB thermodynamic
temperatures $T_i(p)$ in each pixel $p$ of the $i$ channel maps are modeled as
\begin{equation}
T_i(p) = a_iy(p) + b_is(p) + n_i(p),
\end{equation}
where $y(p)$ is the tSZ Compton-y, $s(p)$ is the blackbody primary CMB, and $n(p)$ is noise.
A tSZ Compton-y map is generated by computing a set of weights $w_i$ that recovers the Compton-y, such that 
$y(p) = w_iT_i(p)$. These weights are computed by minimizing the variance in the CMB temperature maps and 
the assumed emission model, while being constrained to produce zero response to the primary CMB emission and 
unit response to a tSZ spectral distortion. The explicit zero response to primary CMB emission in this ILC algorithm
reduces the variance in the resulting stacked maps in comparison to stacking of the raw frequency maps, 
which aids in recovery of the low signal-to-noise tSZ effect. We note that the assumed emission model 
does not include dust emission, which we will discuss in more detail in Section 2.4. The channel maps used in 
this procedure and the final Compton-y map we utilize in our stacking are in HEALpix\footnote{http://healpix.sourceforge.net/} format at a resolution of $N_{\rm{side}}$ = 2048, all smoothed to a common Gaussian 
beam with 10$\arcmin$ FWHM.

	HS14's primary departure from the well-known ILC algorithm in constructing the final tSZ Compton-y
map lies in the variance minimization of linear combinations of the individual channel maps. Instead of a 
minimization based on the auto-spectra of the channel maps, HS14 uses cross-spectra of independent channel 
maps from \emph{Planck}'s single-surveys (`survey 1' and `survey 2') in the variance minimization. 
Furthermore, the linear combination variance is minimized over a restricted multipole range of $300<\ell<1000$, 
down to scales of approximately the beam FWHM. This restriction on the variance minimization focuses the 
minimization on scales of the cosmic infrared background and unresolved IR sources rather than Galactic dust, 
which is heavily masked before minimization (see below). The final Compton-y map we use is an inverse 
variance-weighted co-addition of the two single-survey Compton-y maps.

\begin{figure}[t]
\centering
\includegraphics[width=0.49\textwidth]{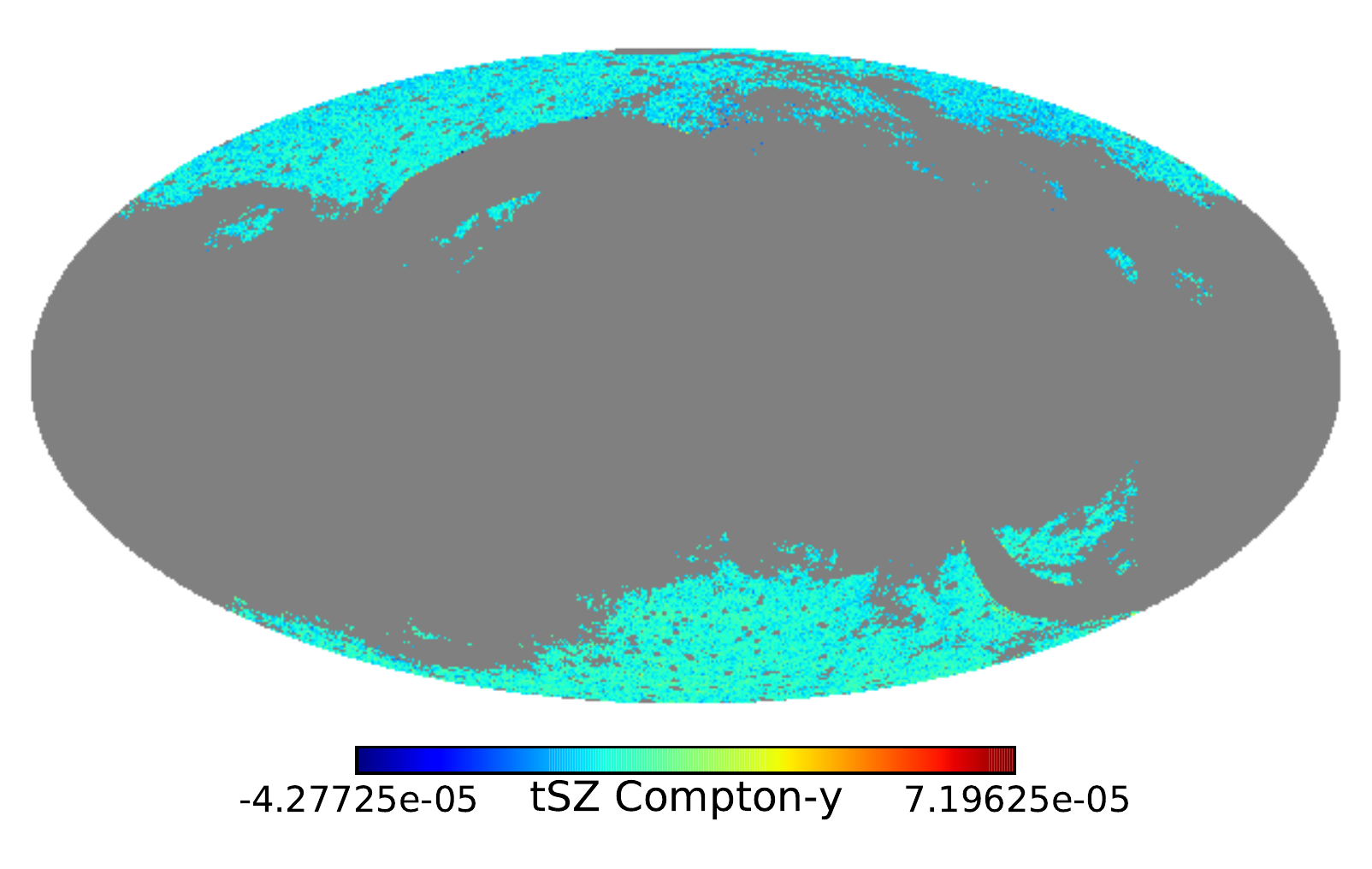}
\includegraphics[width=0.49\textwidth]{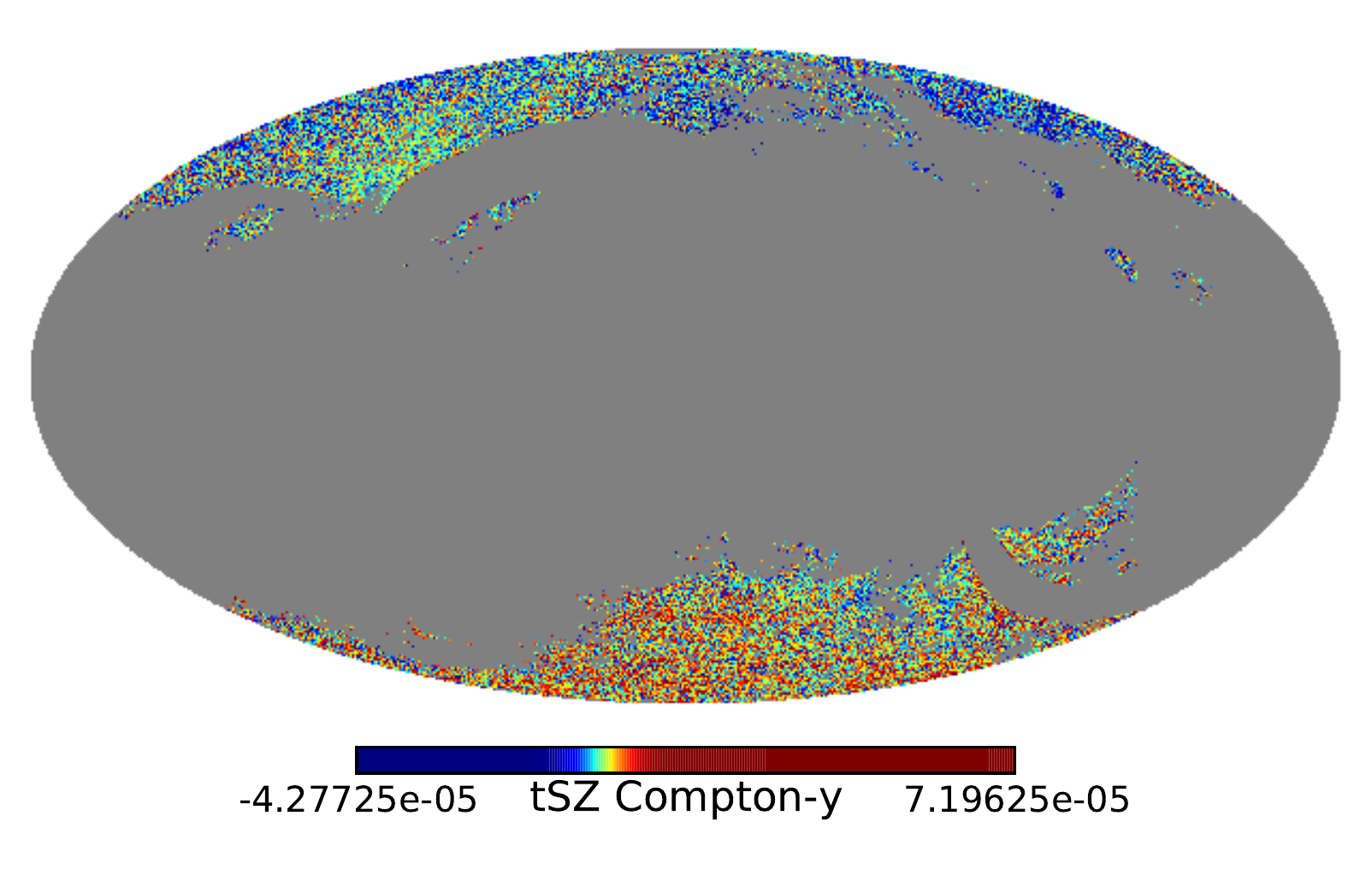}
\includegraphics[width=0.49\textwidth]{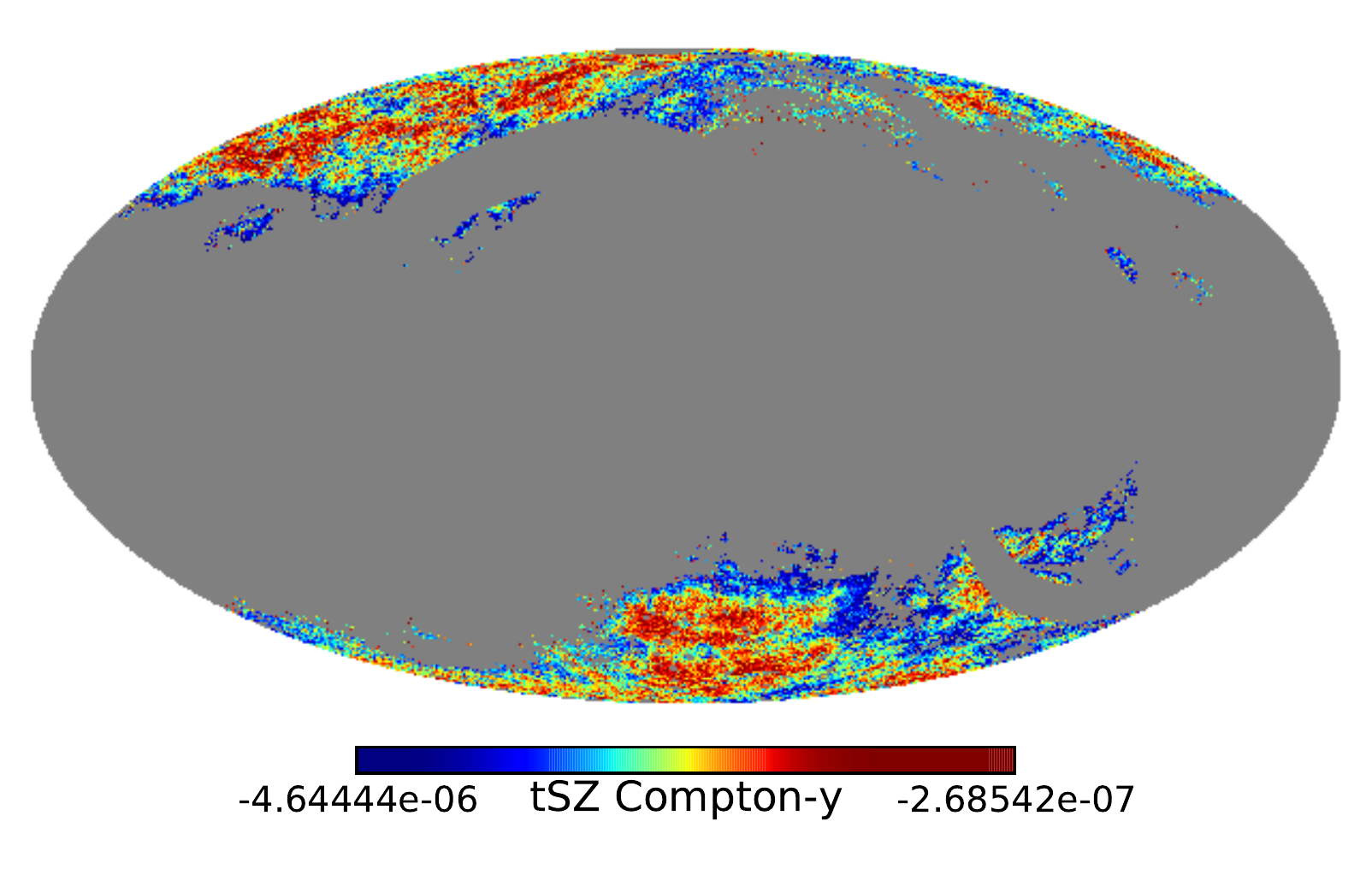}
\caption{
Mollweide projection of the tSZ Compton-y map (mean-subtracted for visualization) of \citet{hi14}, using a linear color bar (top), and a high-contrast histogram-normalized color bar (middle), as well as the expected dust Compton-y contamination map (bottom). The tSZ Compton-y map used in our stacking is corrected for dust contamination by subtracting this dust Compton-y contamination map (see Section 2.4). Various masks described in Section 2.3, including a dust mask based on the \emph{Planck} 857 GHz channel map and a bright tSZ clusters mask, have been applied.}
\label{fig:tSZmaps}
\end{figure}

	Before construction of the Compton-y map, HS14 applied several masks to each of the channel 
maps, including those to account for the incomplete sky coverage of the single-survey maps, 
all 5$\sigma$ point sources detected in any frequency band in both the Low and High Frequency 
Instruments (which removes all resolved radio sources), and regions with significant Galactic dust 
emission. Massive galaxy clusters are the brightest individual sources in the final tSZ map of 
HS14, but their presence in our Compton-y map stacking procedure (see Section 3) can strongly increase 
the variance of the background in our mean stacked maps, which instead aim to detect tSZ signal
from halos of much lower masses and tSZ Compton-y's. We thus apply an additional bright tSZ cluster 
mask to the final Compton-y map of HS14, based on the \emph{Planck} Early Sunyaev-Zel'dovich Detected 
Cluster Candidates Catalog \citep{planck11}; all pixels within the cluster angular extent listed in the catalog based 
on X-ray or \emph{Planck} observations (where the X-ray extent is not available) are masked.

	To minimize dust contamination in their Compton-y map, HS14 preemptively masked regions 
of strong dust emission in the channel maps by removing all HEALpix pixels with intensity above the 
30th percentile in the \emph{Planck} 857 GHz zodi-corrected channel map. However, HS14 show that 
the distribution of the tSZ Compton-y in all pixels in their map shows a non-Gaussian tail towards negative 
Compton-y's, indicating that some residual Galactic dust contamination still persists. We thus create an 
additional dust mask using the same method of HS14, but instead placing the 857 GHz intensity threshold 
to remove all pixels above the 20th percentile. In our tests, forgoing this additional dust mask and the 
bright tSZ cluster mask increases the variance in the background of our stacked Compton-y masks,
thus reducing the signal-to-noise of our detections in Section 3. The union of our final mask and 
the mask of HS12 leaves a Compton-y map covering 16.4\% of the sky (in comparison to the 25.2\% 
in the HS14 map), but the map is still well-matched to the footprint of the SDSS quasar and galaxy sample 
that we will use. Figure 1 (top panel) shows the the resultant masked Compton-y map (mean-subtracted 
for visualization) used in this investigation.

\subsection{Corrections for Dust Contamination}
	The primary drawback of using the tSZ Compton-y map of HS14 for our analysis is that it does not 
include a dust component in the emission model. When multiplied by the ILC weights, dust emission 
in the five channel maps used in the ILC algorithm can produce a non-zero response, allowing an additional `false' 
Compton-y signal to leak into the final Compton-y map. Even after imposing our more stringent Galactic dust mask, 
the Compton-y map of HS14 is not completely free of dust contamination. We find that the tSZ Compton-y signal reconstruction in HS14's modified ILC algorithm produces a negative response to dust emission, primarily due to 
the negative weighting to the high-frequency 545 GHz channel that is produced by the ILC variance minimization. 
In other words, the ILC weight $w_i$ that is multiplied by the thermodynamic temperature in the 545 GHz 
channel map is negative, which can cause the final Compton-y to be under-predicted for strong dust emission in 
this high-frequency band. This can result in systematically under-predicted or even negative tSZ Compton-y's in 
regions of strong dust emission, especially for dust emission with intensities that increase with frequency. 
These effects on the resulting Compton-y maps are evident in the negative Compton-y regions at low Galactic 
latitudes in Figure 1 (top panel), and more clearly seen when the color contrast is changed from linear to histogram-normalized in Figure 1 (middle panel).

	Figure 1 (middle panel) also reveals large-scale fluctuations in the Compton-y background, 
which shows systematic differences in the mean Compton-y across large regions (e.g., the south Galactic
cap has systematically larger Compton-y than the north). These fluctuations are in part
due to the limited multipole range of $300<\ell<1000$ in the variance minimization of HS14's modified ILC 
algorithm. This limited multipole range in the variance minimization allows dust emission on large scales to produce systematic fluctuations in the Compton-y map at lower multipoles, which are primarily driven by large-scale 
Galactic dust emission. This issue is also evident in the auto-power spectrum of the original Compton-y map
(see Figure 9 of HS14), where the power on scales $\ell\lesssim300$ is systematically higher than that
found by \citet{planck13b}. To mitigate the effects of this non-uniform background in our analysis, 
we will subtract the local background in a 20-30$\arcmin$ annulus around each object in our stacking 
procedure (see Section 3). Although this background normalization in our stacking minimizes the effects 
of the large-scale Compton-y fluctuations, it does not fully take into account the negative Compton-y response 
in the ILC algorithm for \emph{intrinsic} dust emission from quasars and galaxies. This is especially relevant for our 
investigation since strong negative k-corrections of quasar and galaxy dust emission in the millimeter/sub-mm 
spectrum can cause the observed dust emission flux densities to be nearly redshift-independent.

	To explicitly correct for negative Compton-y response in the ILC algorithm from dust emission intrinsic to 
galaxies and quasars, we utilize the 857 GHz \emph{Planck} zodi-corrected nominal channel map (smoothed 
to the common 10\arcmin~FWHM) as a tracer for the expected dust emission in the other channels utilized in 
HS14's modified ILC algorithm. We assume a fiducial single-temperature modified blackbody for the dust with 
spectrum $f_\nu \propto B_\nu(\nu)\nu^\beta$  \citep[e.g.][]{bl03}, where $\beta = 2$. The dust temperature in 
this fiducial spectrum is fixed to $T_\mathrm{dust} = 20$ K and 34 K in the rest-frame for our galaxy and quasar 
stacks, respectively, consistent with inferred dust temperatures in star-forming galaxies \citep[e.g.][]{cl13} and 
quasars at $z \simeq 1.5$ \citep[e.g.][]{da12}. The redshifted dust temperatures $T_\mathrm{dust}/(1+z)$ at the 
mean redshifts of our galaxy and quasar samples are conveniently all $\simeq$13.3 K, and so we adopt this single 
redshifted dust temperature for our fiducial dust spectrum.

	Using the above fiducial dust spectrum with redshifted $T_\mathrm{dust}/(1+z) = 13.3$ K, we extrapolate 
the brightness temperature of dust emission in each of the other HFI channels based on the normalization provided 
by the 857 GHz channel map and the effective frequencies of the HFI channels provided in the Planck Explanatory 
Supplement\footnote{http://www.sciops.esa.int/wikiSI/planckpla}. We then produce a map of the expected 
negative Compton-y response due to dust emission by multiplying the expected dust emission in each channel 
(in thermodynamic temperature units) by the ILC weights of HS14 for the public 30\% threshold map we use 
(see their Table 2). The resulting dust Compton-y map is also shown in Figure 1 (bottom panel), and shows 
systematically negative Compton-y's due to dust contamination. A stacking test of our dust map 
on our quasar sample shows an excess centered on the quasars that corresponds to integrated 
dust luminosities of order 10$^{46}$ erg s$^{-1}$. To correct HS14's tSZ Compton-y map for this dust emission 
intrinsic to galaxies and quasars, we subtract this dust Compton-y map (thus increasing the Compton-y's), 
and the final dust-corrected tSZ Compton-y map is used in our quasar and galaxy stacking procedure (Section 3).

	We empirically tested the sensitivity of our results to the assumed dust spectrum by also producing stacked 
Compton-y maps for our quasar and galaxy samples using various spectral indices $\beta$ from 1.0 to 2.0 (which 
spans the range expected for thermal emission in theoretical dust models; \citealt{drainetextbook}), as well as 
redshifted dust temperatures $T_\mathrm{dust}/(1+z)$ from 5 K to the high-temperature Rayleigh-Jeans tail. The 
resulting difference in the Compton-y's is $\lesssim$10\% and are stable around our fiducial dust spectrum, 
showing that our results are robust to the assumed dust spectrum.

\begin{figure}[t]
\centering
\includegraphics[width=0.46\textwidth]{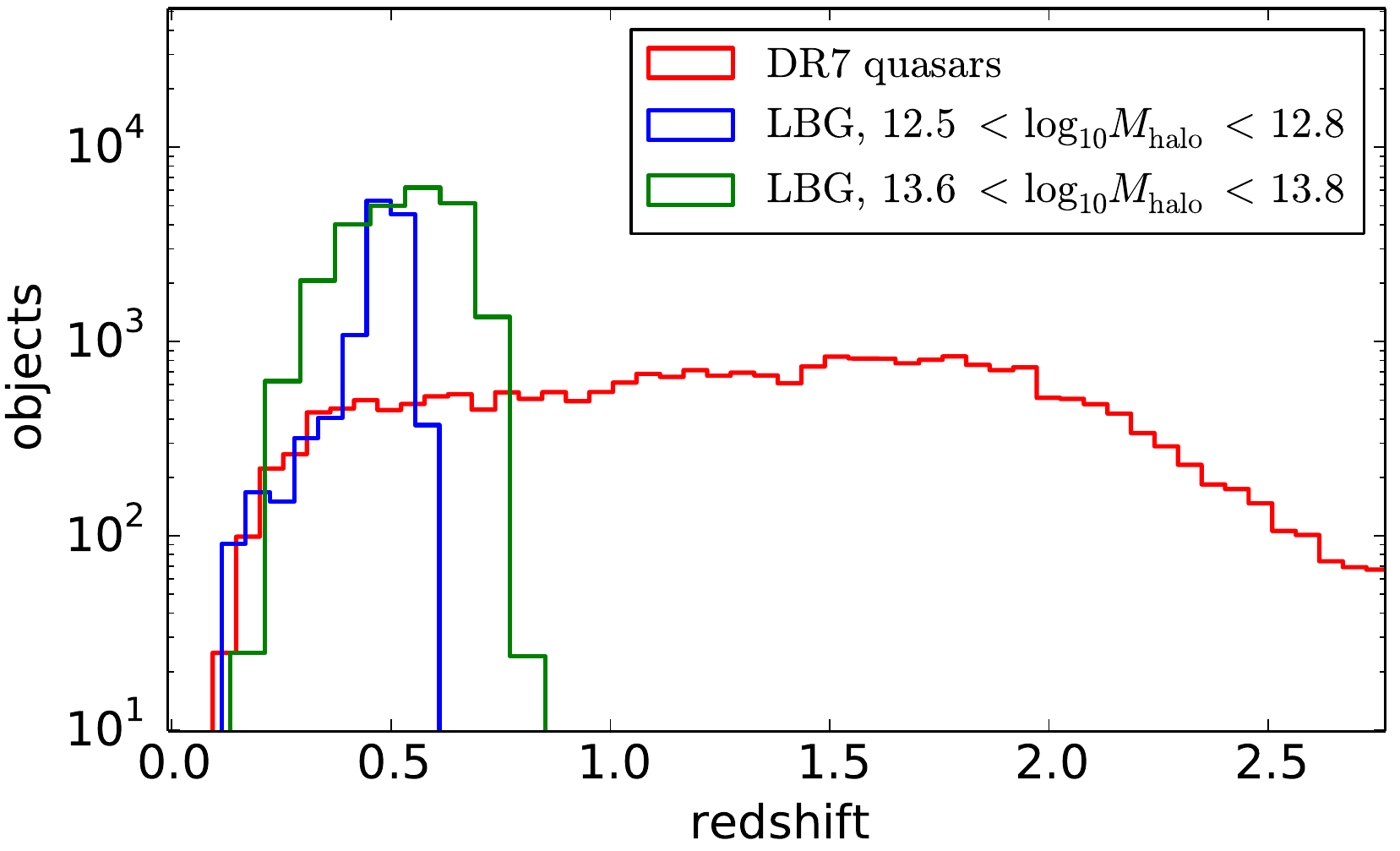}
\caption{
Redshift distributions for our quasar and Locally Brightest Galaxy samples. The median redshifts of the quasar, lower- and higher-mass LBG samples are $z = 1.5$, $0.49$, and 0.54, respectively.
}
\label{fig:redshift_distributions}
\end{figure}
	
\subsection{SDSS Spectroscopic Quasar Sample}
	We use a sample of spectroscopically identified quasars from the Sloan Digital Sky Survey 
\citep[SDSS,][]{yo00} Data Release 7 (DR7) Quasar Catalog \citep{sc07}. We also utilize various 
additional properties of these DR7 quasars calculated and complied in \citet{sh07}, including
virial black hole masses $M_{\rm{BH}}$, bolometric luminosities $L_{\rm{bol}}$, and radio
flux densities. This DR7 quasar catalog contains 105,783 quasars, selected primarily 
by optical color \citep{ri02}. The spectroscopic footprint covers approximately 10,000 deg$^2$ 
over the north and south Galactic caps, largely overlapping the \emph{Planck} tSZ 
Compton-y map. From this DR7 catalog, we select 39,117 quasars which fall in HEALpix 
pixels not masked in the various tSZ Compton-y map masks described in Section 2.3.

	The tSZ effect at millimeter wavelengths from quasars may be susceptible to contamination 
from intrinsic radio emission, particularly since approximately 10\% of spectroscopic quasars 
in SDSS are radio-loud \citep{iv02, ba12}. Although the \emph{Planck} point sources mask imposed on the 
Compton-y map removes all resolved radio sources detected in any \emph{Planck} frequency channel, 
some unresolved radio emission may still be present. We thus remove all quasars in our sample 
with observed radio flux density at rest-frame 6 cm above 2 mJy observed by the FIRST radio survey 
\citep{be95, wi97}, as well as a small number outside the FIRST survey footprint for which the radio 
flux densities are unconstrained (in total $\simeq$10\% of our sample). A separate stacking test of 
our quasar sample that includes the radio-bright quasars in our sample produces only slightly lower
mean Compton-y's.  Furthermore, we find that even if all of our radio-quiet quasar sample had fluxes 
just below the FIRST point source threshold, they would contribute just $20\%$ to the $Y$ we detect 
(assuming a mean spectral index in specific intensity of $-0.7$; \citealt{iv02}).  Thus, we conclude that radio emission does not significantly contaminate our stacks.  

\begin{figure}[t]
\centering
\includegraphics[width=0.47\textwidth]{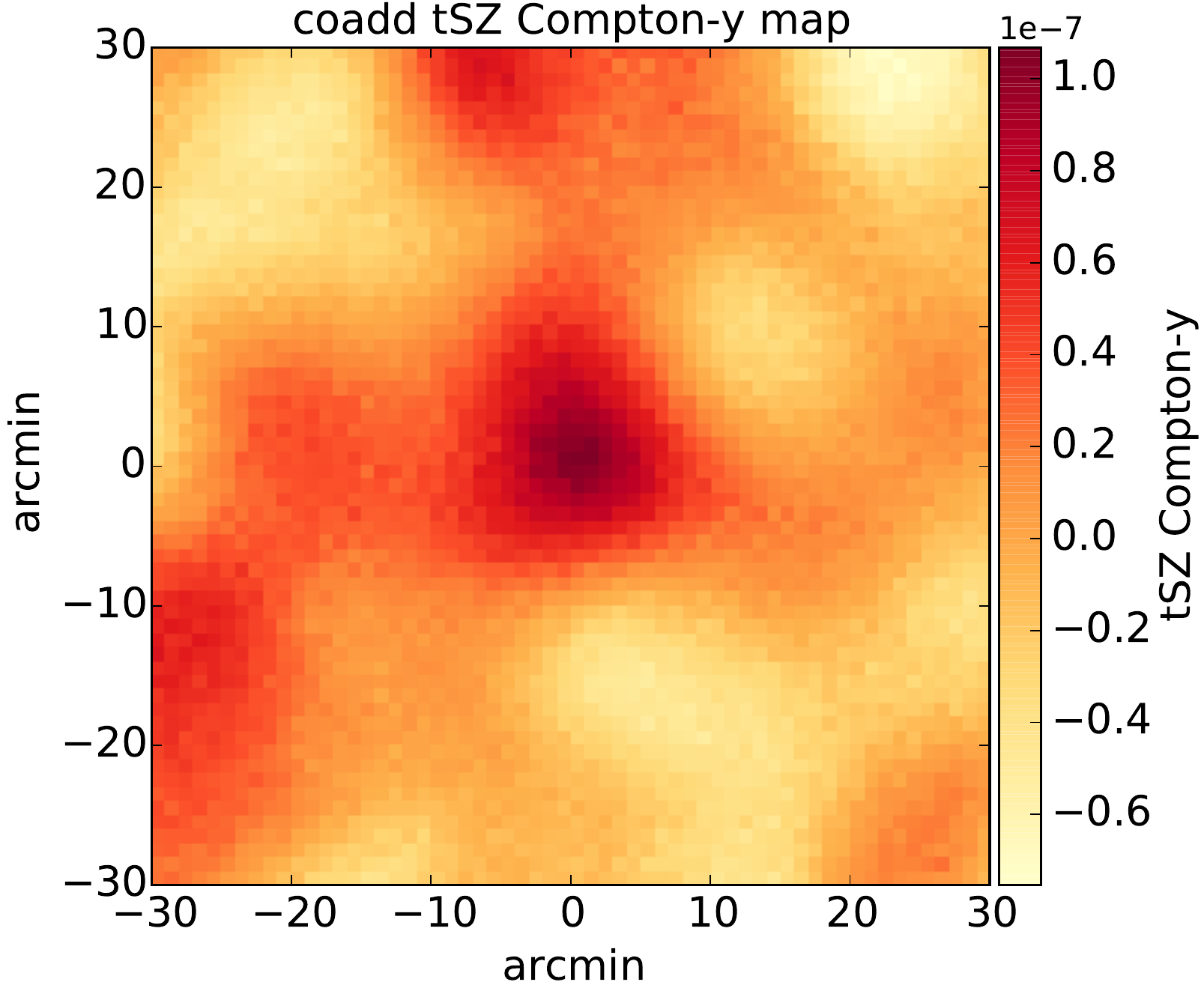}
\caption{Mean tSZ map from our stack of 26,686 quasars. A clear tSZ signal centered on the quasars is evident. 
The coherence length of structures in this stacked map is set by the 10\arcmin~FWHM Gaussian beam,
and the quasar tSZ signal is unresolved.
}
\label{fig:coadd_map}
\end{figure}

\subsection{SDSS Spectroscopic Locally Brightest Galaxy Sample}
	To verify that the mean quasar tSZ signal measured from our Compton-y map stacking procedure
is robust, we also stack samples of galaxies and compare their tSZ signal to those measured by PC13a.
Specifically, we will use SDSS spectroscopic galaxies and construct samples of Locally Brightest Galaxies 
(LBGs), following the procedure of PC13a. These LBGs are the most luminous in their local vicinity, and are 
thus likely the main galaxy in their halos. The unobservable halo mass of each LBG in our sample can be estimated 
using estimates of their stellar mass, along with the stellar-to-halo mass (SHM) relation calculated 
by PC13a. This SHM relation is based on an updated version of the semi-analytic galaxy formation 
simulation of \citet{gu11}, which PC13a used to select a simulated LBG sample, and then calculated a 
SHM relation based on their stellar and halo masses in the simulation. 

	We use the SDSS Data Release 10 \citep[DR10,][]{ah14} spectroscopic galaxy sample, which includes
stellar mass estimates from the Portsmouth Group based on their SDSS $ugriz$ photometry. These galaxy 
stellar masses are estimated using the method of \citet{ma06} and the stellar populations model of \citet{ma09}. 
We select spectroscopic galaxies with observed $r_{\mathrm{AB}} < 21$ model mag from the photometry, 
and spectroscopic redshifts $z > 0.03$. To remove satellite galaxies, we select samples of LBGs following 
the method of PC13a, in which the SDSS photometric galaxy catalog is used to remove spectroscopic galaxies 
with brighter nearby photometric companion galaxies. Specifically, we remove any spectroscopic galaxy that has 
a nearby photometric galaxy with brighter $r$ model mag, and is within a projected distance of 1.5 $h^{-1}$ Mpc 
and redshift separation $<$2000 km s$^{-1}$. The redshifts of the photometric galaxies used in this procedure 
are photometric redshifts, estimated using the kd-tree nearest neighbor fit described in \citet{cs03} and provided 
as part of SDSS DR10. This selection of LBGs allows us to directly use the SHM relation of PC13a to estimate 
their halo masses, and it was shown in PC13a to be robust against changes in the projected distance parameters 
of 1.5 $h^{-1}$ Mpc and 2000 km s$^{-1}$, with low contamination ($\lesssim$ a few percent). 

	We first construct a lower-mass LBG sample with halo masses similar to the mean quasar
$M_\mathrm{halo} = 10^{12.7}$ $M_\odot$; these are approximately the lowest mass galaxies 
for which tSZ signal is still marginally detected by PC13a. Specifically, we select a 
sample of 70,889 LBGs with stellar masses $10^{10.9} < M_\star < 10^{11.1}$ $M_\odot$ 
that are not in masked pixels of our tSZ Compton-y map. LBGs with these stellar masses lie in halos 
with mass approximately in the range $10^{12.5} < M_{\mathrm{halo}} <  10^{12.8}$ $M_\odot$.  We note 
that halo masses provided by the SHM relation of PC13a are presented in units of $M_{500}$, while the 
observed quasar halo masses estimated from clustering measurements we compare to are generally in 
units of $M_{200}$. For Navarro-Frenk-White halos \citep{na06} of mass $M_{200} = 10^{12.7}$ $M_{\odot}$ 
(similar to quasars), and a concentration parameter consistent with that found by \citet{ne07}, $M_{500}$ is 
$10^{12.6}$ $M_{\odot}$. Since $M_{500}$ and $M_{200}$ are not significantly different at the halo mass 
scales we probe, we treat them as equivalent quantities and refer to them as $M_{\mathrm{halo}}$.

	We also construct a sample of higher-mass LBGs that have stronger tSZ signal, for 
comparison to PC13a. We select 81,766 galaxies with stellar mass 
$10^{11.4} < M_{\star} <  10^{11.6}$ $M_\odot$ (corresponding to halo masses of 
$10^{13.6} < M_{\mathrm{halo}} < 10^{13.8}$ $M_\odot$). The details of these two LBG samples 
are also summarized in Table 1, and their redshift distributions are also shown in Figure 2, with median 
redshift $z = 0.49$ and $0.54$ for the lower- and higher-mass LBG samples, respectively. 

\section{Stacking Compton-y Maps of the Quasar and Galaxy Samples}	
\label{sec:stack}
	To detect tSZ signal from our quasar and Locally Brightest Galaxies (LBGs) samples,
we will stack regions of the Compton-y map described in Section 2.3, centered on each of the different 
samples. We will first do this in angular space for an initial detection, and then in physical space for the
actual measurement of the mean integrated Compton-y's. While PC13a also showed stacks of the tSZ 
(constructed using an ILC algorithm) centered on their LBG galaxy samples, their actual tSZ measurements 
were made using a multi-frequency matched filter (MMF) algorithm. The MMF algorithm used by PC13a assumes 
a template tSZ profile for the sources, perhaps better-suited for the tSZ signal from galaxies where there is some 
understanding of the pressure profile. However, the pressure profiles of ionized gas in quasar host halos due to 
feedback is unconstrained. Therefore, it makes sense to stack around the positions of the quasars, and 
apply an aperture photometry approach to empirically measure the integrated Compton-y's without assuming a 
radial template. We will verify that this approach on our LBG samples yields a signal that is consistent with 
those reported in PC13a.

\subsection{Angular-Space Stacking}
	For each quasar in our sample, we cut out a 1$^{\circ}$ by 1$^{\circ}$ square region 
from the masked tSZ Compton-y map, projected in a gnomonic projection and centered on the quasar. 
Each of these regions is comprised of 60$\times$60 pixels at resolution 1$\arcmin$/pixel, and their rotational 
orientation are chosen at random. We will not include any quasars in our stack for which 
more than 10$\%$ of pixels within 30$\arcmin$ of the central quasar are masked. To mitigate the effects 
of large-scale fluctuations in the background of the tSZ Compton-y map noted in Section 2.4, 
we subtract the mean Compton-y in a background annulus of 20-30$\arcmin$ around each quasar. We produce an 
equal-weight mean co-added map of the remaining 26,686 quasars in our sample, which is shown in Figure 3
and has an implicit 10\arcmin~FWHM Gaussian beam from the original Compton-y map. A clear excess is 
present in this stacked map peaking at the location of the quasars. The above process is repeated to also 
create separate co-additions of our LBG samples.

\begin{deluxetable*}{cccccc}
\tablecolumns{12}
\tablewidth{0pt}
\tablecaption{integrated Compton-y values for our samples of quasars and locally brightest galaxies}
\tablehead{ \colhead{Sample} & \colhead{Halo Mass} & \colhead{Median} & \colhead{$\tilde{Y}^\textrm{a}$}  & \colhead{$Y^\textrm{b}$} & 
\colhead{$E_\mathrm{tot}^\textrm{c}$} \\
& \colhead{[$M_\odot$]} & \colhead{Redshift} & \colhead{[10$^{-6}$ arcmin$^2$]} & \colhead{[10$^{-6}$ Mpc$^2$]} & \colhead{[$10^{61}$ erg]}
}
\startdata
SDSS DR7 Quasars  &  median $\mathrm{log}_{10}M_{\mathrm{halo}}  \simeq 12.7$ & 1.50 & 96 $\pm$ 19 & 3.5 $\pm$ 0.7  & 11 $\pm$ 2.2 \\
\rule{0pt}{2.5ex}  

SDSS DR7 Quasars, higher-$z$  &  median $\mathrm{log}_{10}M_{\mathrm{halo}}  \simeq 12.7$ & 1.96 & 115 $\pm$ 19 & 4.8 $\pm$ 0.8  & 15 $\pm$ 2.8 \\
SDSS DR7 Quasars, lower-$z$  &  median $\mathrm{log}_{10}M_{\mathrm{halo}}  \simeq 12.7$ & 0.96 & 74 $\pm$ 30 & 2.2 $\pm$ 0.9  & 7.1 $\pm$ 3.1 \\
\rule{0pt}{2.5ex}  

LBG, $10.9 < \mathrm{log}_{10}M_\star < 11.1$ & $12.5 < \mathrm{log}_{10}M_{\mathrm{halo}} < 12.8$ $M_\odot$ & 0.49 & 4 $\pm$ 20 & 0.1 $\pm$ 0.5 & 0.3 $\pm$ 1.6 \\
LBG, $11.4 < \mathrm{log}_{10}M_\star < 11.6$ & $13.6 < \mathrm{log}_{10}M_{\mathrm{halo}} < 13.8$ $M_\odot$ & 0.54 & 56 $\pm$ 16 & 1.4 $\pm$ 0.4 & 4.5 $\pm$ 1.3 \\
\rule{0pt}{2.5ex}  

virialized quasar halo gas, 5-param HOD & & 0.5 & 42 & 1.0$^\textrm{d}$ & 3.2$^\textrm{d}$  \\
virialized quasar halo gas, 6-param HOD & & 0.5 & 27 & 0.7$^\textrm{d}$  & 2.3$^\textrm{d}$ 
\enddata
\tablenotetext{a}{Redshift-independent integrated Compton-y, Equation 6} 
\tablenotetext{b}{Integrated Compton-y, Equation 3} 
\tablenotetext{b}{Total thermal energy in ionized gas, Equation 5} 
\tablenotetext{d}{These $Y$ and $E_\mathrm{tot}$ values for virialized halo gas in quasar host halos at $z=0.5$ have been rescaled to the $z=0.96$ median redshift of our low-$z$ quasar sample for direct comparison. See Section 4.1 for details.} 
\end{deluxetable*}

	Figure 4 shows the radial profile of the tSZ Compton-y signal from the stacked image of our quasar 
sample, beginning from the center of the image in Figure 3, using bins of width 1.25$\arcmin$.
Figure 4 also shows the Compton-y radial profile of a stacked quasar map without our dust corrections 
for comparison; as expected, correcting for the negative ILC Compton-y response to intrinsic dust emission 
increases the tSZ signal. We perform a null test of our detected quasar tSZ signal by stacking on the locations of 
$2\times10^6$ randomly-distributed points on the Compton-y map for which $<$10\% of the pixels in the stacked
region are masked. The Compton-y profile of this stacked map is also shown in Figure 4, and is consistent
with a non-detection. A separate null test (not shown) in which the position of each quasar in our sample is 
randomly offset by 1$^{\circ}$ in Galactic latitude before stacking also shows a similar non-detection.

\begin{figure}[t]
\centering
\includegraphics[width=0.48\textwidth]{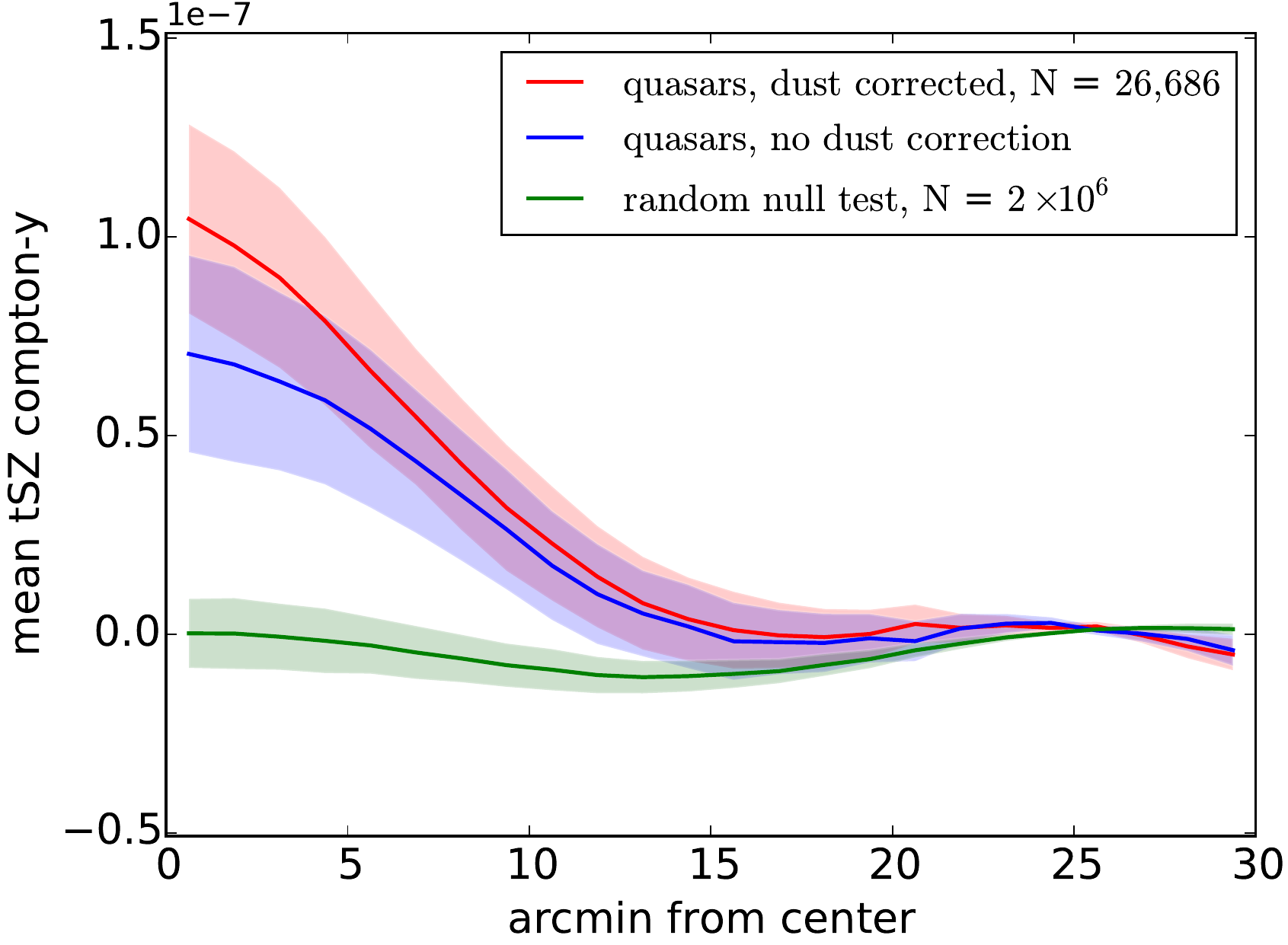} 
\caption{
Radial Compton-y profile of the mean stacked map centered on the quasar sample with correction for dust 
contamination (red line), and without dust corrections (blue line), as well as a null test (green line).
Dust contamination in the ILC algorithm produces a negative Compton-y response that we find is robust to 
the dust spectral template, and thus the quasar tSZ signal is stronger after dust corrections. The 1$\sigma$ 
uncertainties shown (shaded regions) are estimated using a bootstrap resampling of the quasar-centered cutouts 
comprising each stacked map.  
}
\label{fig:tSZ_profile_quasars}
\end{figure}

	We produce separate stacked tSZ images of our two LBG samples (summarized in Table 1), 
and their mean Compton-y radial profiles are shown in Figure 5, along with the dust-corrected quasar 
radial profile from Figure 4 for comparison. The lower-mass LBG sample does not show a significant  
tSZ signal, while our higher-mass LBG sample shows a clear detection; these results are qualitatively consistent
with those of PC13s for LBGs in these mass ranges, and we make a quantitative comparison of their integrated
Compton-y's in Section 4.1. We note that the samples of quasars and LBGs used in the stacked Compton-y maps 
in Figures 4 and 5 are smaller than the original samples defined in Sections 2.5 and 2.6 due to the requirement 
that $<$10$\%$ of pixels in the stacked regions are unmasked. The 1$\sigma$ uncertainties on the radial profiles 
in Figures 4 and 5 are estimated using a standard bootstrap resampling of the individual regions comprising the 
stacked map. Since the number of pixels in each azimuthal bin increases with radius, the uncertainties in the 
Compton-y radial profiles are naturally smaller at large radii. 

\begin{figure}[t]
\centering
\includegraphics[width=0.48\textwidth]{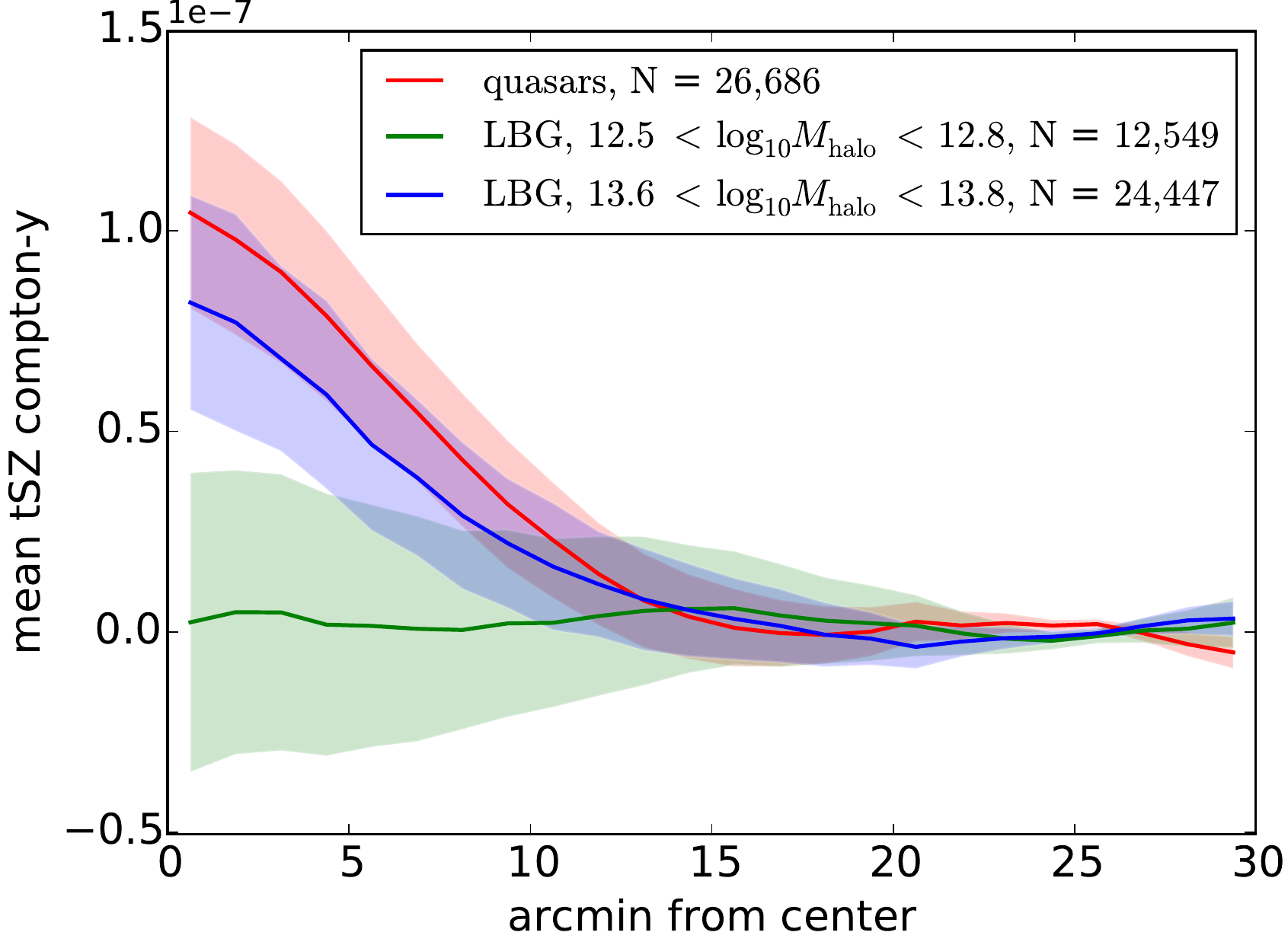}
\caption{
Radial Compton-y profile of mean stacked maps (all dust corrected) centered on the quasar sample (red line), 
the lower-mass LBG galaxy sample (green line), and higher-mass LBG sample (blue line). The mean integrated Compton-y of our LBG samples are consistent with the results of PC13a, indicative of hot halo atmospheres in these galaxies. However, the Compton-Y of the quasar sample is significantly larger than that expected from halos of similar masses (compare green and red curves), 
suggesting additional heating from quasar feedback (see Section 4.1).
}
\label{fig:tSZ_profile_galaxies}
\end{figure}

\subsection{Physical-Space Stacking}
	Due to the different angular diameter distances of our Locally Brightest Galaxies (LBGs) and quasar samples 
(as well between objects in each sample), their mean integrated Compton-y (Equation 3) and thermal energy in 
electrons (Equation 4) should be measured and compared in physical space (although we note that the angular 
diameter distance only varies by approximately 50$\%$ between $z=0.5$ and $z=2$, making the angular and physical 
space stacks qualitatively similar). We produce a separate stacked image of the quasar 
sample using regions of 30 $h^{-1}$ Mpc by 30 $h^{-1}$ Mpc in physical distance at the redshift of each 
quasar. At the median quasar redshift of $z=1.5$, this is approximately the same region in physical 
space as the 1$^{\circ}$ by 1$^{\circ}$ regions used in the angular-space stacked image of Figure 3. 
Intentionally scaling the size of the regions in this manner causes the resulting physical-space stacked
image (not shown) to look similar to that in angular-space in Figure 3, and allows us to use a background 
annulus of 10-15 $h^{-1}$ Mpc around each quasar for background subtraction (similar to the 20-30\arcmin~in 
the angular-space stacks). Similarly, for the physical space stacks of our LBG samples, we use regions of 
20 $h^{-1}$ Mpc by 20 $h^{-1}$ Mpc at the redshift of each LBG (also approximately a 1$^{\circ}$ by 1$^{\circ}$ 
region at the LBG median redshift of 0.52), and background annulus of 6.67-10 $h^{-1}$ Mpc.
The similarity in size of the stacked regions in physical and angular spaces also causes the
Compton-y radial profiles of our quasar and LBG samples in physical space to be similar to 
those in angular space (Figures 4 and 5). 

\subsection{Integrated Compton-y Measurements: \\ Comparison to PC13a}

	We calculate the mean $Y$ (Equation 3) and $\tilde{Y}$ (Equation 5) of our quasar 
and LBGs from their median Compton-y radial profiles of our physical space stacks. For the integrated 
Compton-y, we take a simple aperture photometry approach to the radial profiles of the previously 
background-subtracted Compton-y stacked maps by integrating their Compton-y profiles within a central 
circular source region. For the quasar and LBG samples, this source aperture is chosen to be at radius 
5 $h^{-1}$ Mpc and 3.75 $h^{-1}$ Mpc, respectively. These aperture radii in physical space 
correspond to approximately 10$\arcmin$ for the quasar and LBG samples at their median redshifts 
in the angular-space Compton-y profiles in Figure 4, and twice the beam FWHM in the Compton-y map.
	
	The $\tilde{Y}$ (Equation 5), rescaled angularly and in redshift, is estimated based on the resulting 
$Y$ values, by fixing the angular diameter distance to 500 Mpc and assuming the median 
redshifts of $z = 0.52$ and 1.5 for our LBG and quasar samples, respectively. The measured mean
$Y$ and $\tilde{Y}$ values of our quasar and LBG samples are listed in Table 1. The $1\sigma$ 
uncertainties on these integrated Compton-y's are also based on bootstrapping of the individual regions
that comprise their stacked images.

	We verify our tSZ measurements from our stacking procedure by comparing our $\tilde{Y}$
measurements for LBGs to those of PC13a. Comparison of $\tilde{Y}$ 
to previous work should account for any differences in the solid angle over which the Compton-y
is integrated, as well as whether the Compton-y measurements capture the integrated Compton-y in a cylinder
along the line of sight or over an enclosed sphere assuming a pressure profile. Our measurement of $\tilde{Y}$ for 
galaxies using Equation 3 is performed by integrating the line of sight Compton-y over a solid angle corresponding
to the 3.75 $h^{-1}$ Mpc radius of the source aperture; this captures the integrated Compton-y in a cylinder along the 
line of sight. In contrast, PC13a's $\tilde{Y}$ measurement using the MMF approach captures the integrated Compton-y 
inside a spherical volume with mean density of $500\rho_c(z)$, where $\rho_c(z)$ is the critical density of the Universe.
The radius of this volume ($R_{500}$) for galaxies of each halo mass is estimated using the universal pressure 
profile of \citet{ar10}, and we denote this resulting integrated Compton-y reported by PC13a as 
$\tilde{Y}^{sph}_{R500}$.
 
	To account for the differences in source aperture sizes in our comparison of $\tilde{Y}^{sph}_{R500}$ 
of LBGs measured by PC13a to the $\tilde{Y}$ we measure, we calculate the conversion factor 
$\tilde{Y}$/$\tilde{Y}^{sph}_{R500}$ for our galaxy samples. The 3.75 $h^{-1}$ Mpc radius of the circular aperture 
we use to integrate the Compton-y over solid angle for LBGs is approximately 15.8$R_{500}$ and 6.8$R_{500}$ 
for our lower- and higher-mass LBG samples, respectively. By integrating the universal pressure profile of \citet{ar10} 
out to 15.8$R_{500}$ and 6.8$R_{500}$ in a cylinder along the line of sight (to find $\tilde{Y}$), and dividing by the integrated pressure profile in a spherical volume out to $R_{500}$ (to find $\tilde{Y}^{sph}_{R500}$), we find that 
$\tilde{Y}$/$\tilde{Y}^{sph}_{R500} = 1.52$. This value of $\tilde{Y}$/$\tilde{Y}^{sph}_{R500}$ (consistent with that 
found by \citealt{greco14}) is the same for both our lower- and higher-mass LBG samples despite the different 
aperture sizes we use of 15.8$R_{500}$ and 6.8$R_{500}$, respectively, because the gas pressure in the 
universal pressure profile is negligible beyond 5$R_{500}$.

	For our lower-mass LBG sample (with $10^{10.9} < M_\star < 10^{11.1}$ $M_\odot$), 
our measured $\tilde{Y} = 4 \,(\pm 20) \times 10^{-6}$ arcmin$^2$ is consistent with the 
 $\tilde{Y} = (1.9 \pm 1.2 ~\mathrm{to}~2.3 \pm 0.9) \times 10^{-6}$ arcmin$^2$ detected at low significance 
 by PC13a (see their Table 1, multiplied by $\tilde{Y}$/$\tilde{Y}^{sph}_{R500} = 1.52$), 
as well as the null detection by \citet{greco14}. The two values of $\tilde{Y}$ measured by PC13a that we are 
comparing to is due to their finer binning of LBG samples in stellar mass, where each of our $M_\star$ bins 
encompasses two bins in PC13a. For our higher-mass LBG sample 
(with $10^{11.4} < M_\star < 10^{11.6}$ $M_\odot$), the $\tilde{Y} = 56 \,(\pm 16) \times 10^{-6}$ arcmin$^2$ 
we measure is in excellent agreement with the strong 
$\tilde{Y} = (44 \pm 5.8~\mathrm{to}~91 \pm 10) \times 10^{-6}$ arcmin$^2$ measured by PC13a. 
This recovery of the tSZ signal from LBGs confirms that our stacking method recovers similar tSZ 
signals as PC13a's MMF approach.

\section{A Feedback Origin for the Quasar tSZ Signal}

\subsection{Separating Quasar Feedback \\ and Virialized Halo Gas}
	The quasar tSZ signal we detected in our Compton-y map stacks is likely to be a combination of 
both feedback effects and virialized halo gas in quasar host galaxies. It may seem that the feedback 
effects can be isolated by simply taking the total quasar tSZ signal and subtracting off the small tSZ signal for 
Locally Brightest Galaxies (LBGs) with halo masses similar to the $10^{12.7}$ $M_{\odot}$ mean halo masses 
of quasars. However, a more careful separation of these two components of the total quasar tSZ signal should 
take into account (1) the full distribution of quasar host halo masses, and (2) the different median redshifts 
between our quasar and LBG samples. Specifically, an estimate of the virialized halo component in the 
quasar tSZ signal based on the $10^{12.7}$ $M_{\odot}$ halo mass LBGs is likely to be biased due to a 
small number of quasars lying in cluster-mass halos. The tSZ signal of these handful of quasars lying in 
clusters are dominated by the ICM rather than feedback, and are orders of magnitude larger than the tSZ 
signal of more typical quasars due to the power-law relation between the tSZ integrated Compton-y and halo 
masses (the $Y-M_\mathrm{halo}$ relation). These quasars can thus significantly increase the mean quasar tSZ 
signal in our stacks, and accounting for this effect requires knowledge of the quasar host halo mass 
distribution. Furthermore, the number of cluster-mass quasar host halos is redshift-dependent, and so the quasar 
tSZ signal should be compared to estimates of the virialized halo gas contribution at similar redshifts.

	An approach we take for isolating the quasar feedback tSZ signal is to create higher- and lower-redshift 
quasar subsamples, with which we produce separate stacks to measure the integrated Compton-y. We argue that 
the high-redshift subsample will not contain quasars lying in massive clusters, and thus its 
tSZ signal will be purely due to feedback with little to no contamination from virialized halo gas. In addition, the 
lower-redshift subsample will be used as a verification of the feedback energetics derived from the high-redshift 
subsample. Specifically, we will estimate the virialized halo gas component in our low-redshift quasar tSZ signal, 
based on the quasar halo mass distribution at $z\simeq0.5$ inferred from quasar-galaxy cross-correlation by 
\citet{sh13}. Since the quasar feedback tSZ signal is a measurement of the total thermal feedback energy in 
ionized gas, we will parameterize this quasar feedback energy $E_\mathrm{fb}$ (calculated using 
Equation 5) as a function of the rest-mass energy of the accreted material (i.e. $M_\mathrm{BH}$), such that 
$E_\mathrm{fb} = \eta\epsilon M_\mathrm{BH}c^2$. The key feedback parameter is the observationally degenerate 
product $\eta\epsilon$, where $\eta$ is the radiative efficiency of accretion, and $\epsilon$ is the feedback coupling efficiency of radiation to the thermal energy in ionized gas. The spectroscopically-derived estimates of $M_{\rm{BH}}$ 
we use for each quasar in our sample are single-epoch $M_{\rm{BH}}$ estimates based on the broad emission lines, 
and have possible systematic uncertainties that are currently poorly-constrained \citep{de09, ra11, de13}.

\begin{figure}[t]
\centering
\includegraphics[width=0.49\textwidth]{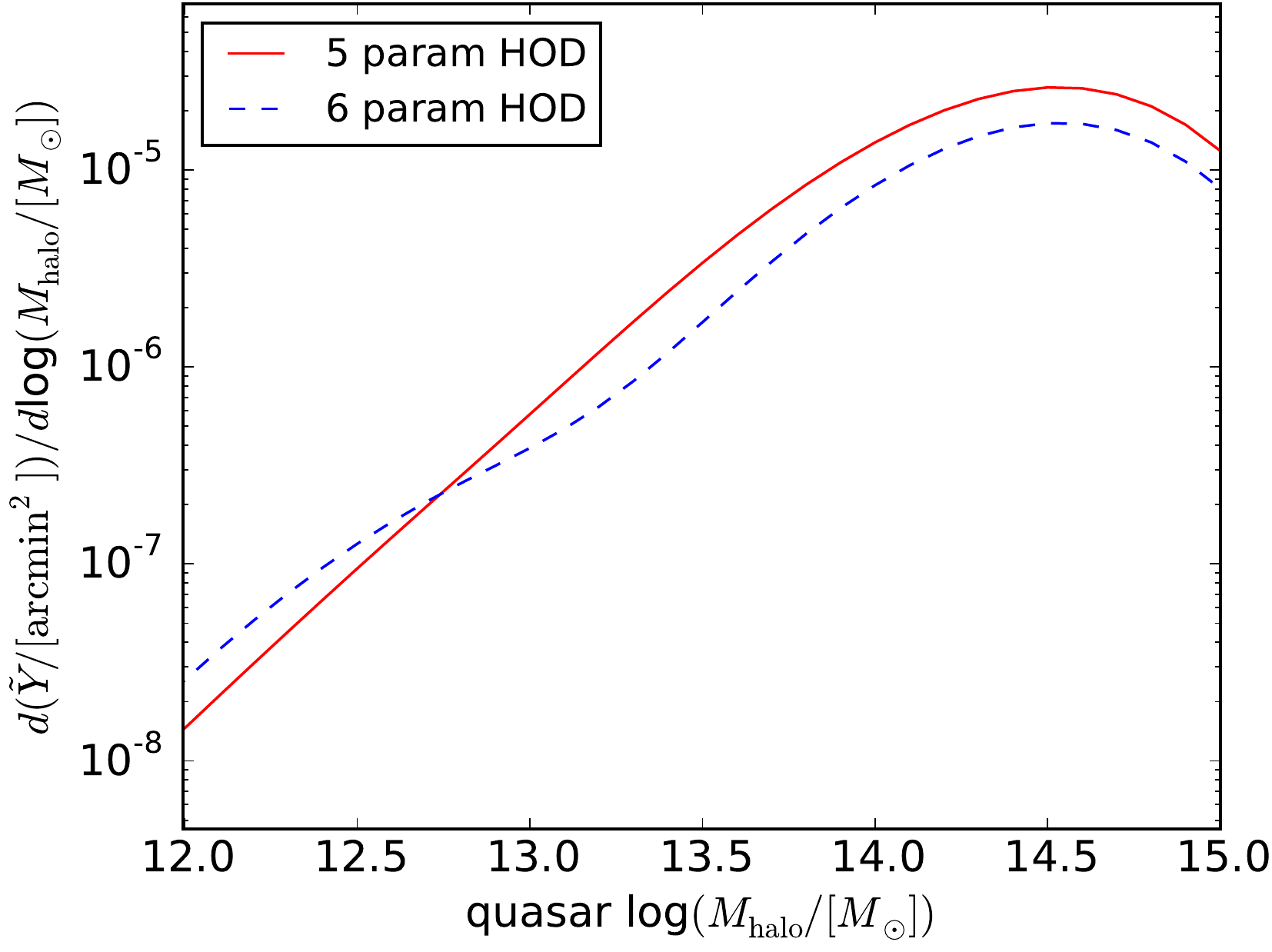} 
\caption{
Contribution to the mean tSZ  $\tilde{Y}$ (Equation 6) from virialized gas in quasar host halos as a function of halo mass, based on the halo mass probability distribution from a 5-parameter (red solid line) and 6-parameter (blue dashed line) halo occupation distribution model.}
\label{fig:tSZ_HOD}
\end{figure}

	To create the high- and low-redshift quasar subsamples, we split our quasars at $z\simeq1.5$,
approximately the median redshift of the main quasar sample. The high- and low-redshift subsamples have
median redshifts of 1.96 and 0.96, and median $M_{\rm{BH}}$ of 10$^{9.2}$ and 10$^{8.8}$ $M_{\odot}$, respectively.
A physical-space Compton-y stack of the high-redshift quasar subsample following the procedure in Section 
3.2 results in an integrated Compton-y of $4.8 \,(\pm0.8) \times 10^{-6}$ Mpc$^2$, implying a total thermal 
energy of $15 \,(\pm2.8) \times 10^{61}$ erg and a quasar feedback efficiency from accretion of  
$\eta\epsilon = 0.05\,(\pm 0.01$ statistical uncertainty), or approximately 5\% of the black hole mass. 
This feedback energy (and the implied feedback efficiency) for high-redshift quasars is not significantly 
affected by virialized halo gas, since the mean tSZ signal for virialized gas in quasar halos at the mean 
quasar halo mass of 10$^{12.7}$ $M_{\odot}$ is negligible (as found by PC13a and \citealt{greco14}, and 
confirmed in our lower-mass LBG stack in Section 3.3). More importantly, the number of quasars lying in cluster-mass 
halos is also effectively zero at $z\simeq1.96$ since they will not have yet formed at these redshifts. 
We note that our quoted uncertainties on $\eta\epsilon$ are statistical only, and do not include the 
poorly-understood systematics in the single-epoch $M_{\rm{BH}}$ estimates we use.

	We further verify this measurement of the quasar feedback efficiency from our high-redshift quasar
sample, by comparing to an analysis of our low-redshift quasar sample. At low-redshifts, the non-negligible tSZ signal from 
virialized halo gas must be subtracted to isolate the quasar feedback energy. To estimate this contribution, 
we utilize the quasar number density as a function of halo mass at $z\simeq0.5$ calculated by \citet{sh13}.
This is inferred from halo occupation distribution (HOD) modeling of the quasar-galaxy two-point cross-correlation 
and when normalized, gives the probability distribution of quasar host halo masses. Different quasar HOD 
models are often known to be degenerate with clustering measurements \citep{ri12, ka12, ch13}, and the 
uncertainties on the best-fit occupation functions are often large. Nevertheless, they provide a useful check of 
our high-redshift feedback results, and so we utilize the quasar $M_{\mathrm{halo}}$ distribution 
inferred in \citet{sh13} using both a 5-parameter and 6-parameter HOD model (see their Figures 14 e and f). 
We multiply these two quasar $M_{\mathrm{halo}}$ probability distributions by the redshift-independent power-law 
relation between $\tilde{Y}^{sph}_{R500}$ and $M_{\mathrm{halo}}$ found in PC13a (see their Equations 1 and 2), 
which results in the contribution to the mean tSZ signal from virialized gas in quasar host halos as a function
of $M_{\mathrm{halo}}$ at $z\simeq0.5$, shown in Figure 6. Since we are comparing 
results from PC13a to our measurements, the integrated Compton-y measurements of PC13a have been 
multiplied by the extra factor of $\tilde{Y}$/$\tilde{Y}^{sph}_{R500} = 1.52$ to account for differences in the 
measurement techniques, as discussed in Section 3.3. Although cluster-sized quasar host halos are rare, 
Figure 6 shows that halos of $\sim$10$^{14.5}$ $M_{\odot}$ are the primary contributors to the mean 
virialized halo gas tSZ signal in quasar host halos at $z\simeq0.5$. Since the virialized halo gas (i.e. ICM) in 
these cluster-mass halos are generally well-modeled as isothermal gas in hydrostatic equilibrium, our direct
comparison of the virialized gas contribution in LBGs to quasars with similar halo masses should not
be affected by additional sources of tSZ signal (e.g. stellar feedback and non-thermal pressure support).

	Integrating the $\tilde{Y}$ contributions in Figure 6 over all halo masses gives the mean $\tilde{Y}$ from
virialized gas in quasar host halos at $z\simeq0.5$; for the 5- and 6-parameter models, we find $\tilde{Y} = 4.2 \times 
10^{-5}$ arcmin$^2$ and $\tilde{Y} = 2.7 \times 10^{-5}$ arcmin$^2$, respectively. Since we will be comparing 
this tSZ signal at $z\simeq0.5$ to our low-redshift quasars at $z\simeq0.96$, the redshift evolution of this virialized halo 
gas tSZ signal and the difference in angular diameter distances should be taken into account. We thus
rescale these mean virialized halo gas $\tilde{Y}$ values to $Y(z=0.96)$ to find $Y(z=0.96) = 1.0\times10^{-6}$ 
Mpc$^2$ and $Y(z=0.96) = 0.7\times10^{-6}$ Mpc$^2$ for the 5- and 6-parameter HOD models, also summarized 
in Table 1. These expected mean quasar halo gas tSZ $Y(z=0.96)$ are smaller than our observed low-redshift 
quasar $Y = 2.2\,(\pm0.9)\times10^{-6}$ Mpc$^2$ by approximately 2 to 3 times for the 5- and 6-parameter HOD 
models, respectively. This suggests the additional presence of feedback effects in our low-redshift quasar sample, 
which likely dominates our observed quasar tSZ signal over the contribution from virialized halo gas.

	To summarize our method above, we have taken the quasar halo mass distribution at $z=0.5$, 
multiplied by the $\tilde{Y}$- $M_{\mathrm{halo}}$ relation for LBGs, and integrated to find the virialized halo 
gas contribution in our low-redshift quasar tSZ $\tilde{Y}$ signal. Since the low-redshift quasars we wish to 
compare to have median redshift of $z=0.96$, we rescaled the estimated virialized halo gas $\tilde{Y}$ (which is 
redshift-independent) to the redshift-dependent $Y(z=0.96)$. However, we note that despite this 
rescaling of $\tilde{Y}$ to $Y(z=0.96)$, $Y(z=0.96)$ is still not a completely accurate estimate
of the virialized halo gas tSZ signal for quasars at $z=0.96$, because the original quasar halo mass distribution
we use is instead measured for $z=0.5$. This will lead us to slightly overestimate the virialized halo gas tSZ 
$Y(z=0.96)$ by overestimating the number of cluster-sized quasar host halos at $z\simeq0.96$, and thus 
underestimate the feedback signal. Limiting our low-redshift quasar sample to only quasars at $z\simeq0.5$ results 
in a subsample that is too small to produce a high signal to noise stacked tSZ detection. Nevertheless, as a useful 
additional check of the high-redshift feedback parameters, we isolate the feedback tSZ signal for our low-redshift 
sample at $z\simeq0.96$ by subtracting our best estimate of the mean virialized halo gas component from the 
5- and 6-parameter HOD models. We find that the isolated low-redshift quasar feedback has mean energies of
$4.4 \,(\pm3.1) \times 10^{61}$ erg, giving a quasar feedback efficiency from accretion of $\eta\epsilon \simeq 0.038$ 
($\pm$ 0.027 statistical uncertainty). Considering that this is likely to be a small underestimate of $\eta\epsilon$, we 
conclude that this is crudely consistent with the $\eta\epsilon = 0.05$ feedback efficiency we measure at high redshifts.

\begin{figure}
\centering
\includegraphics[width=0.47\textwidth]{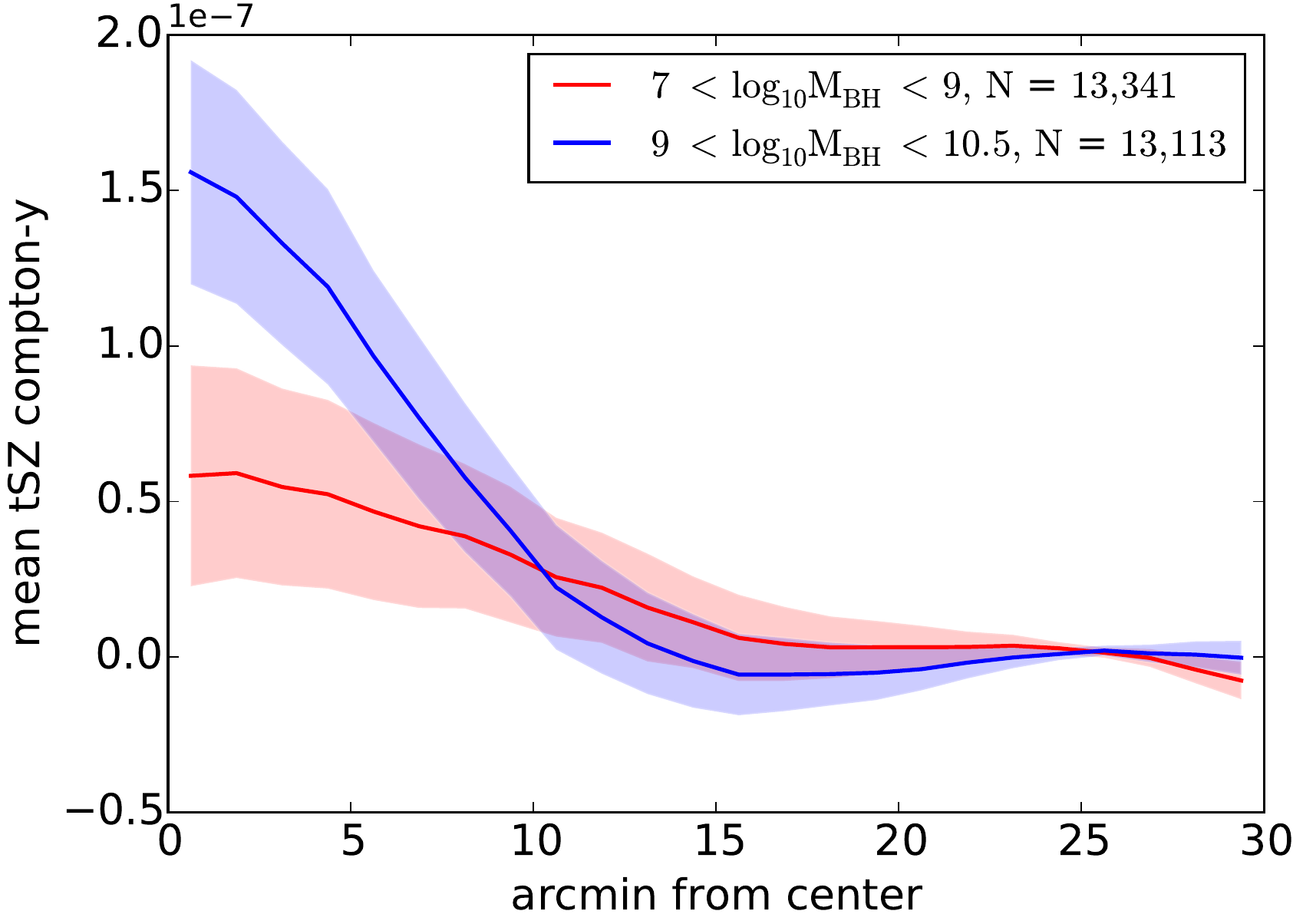} 
\includegraphics[width=0.47\textwidth]{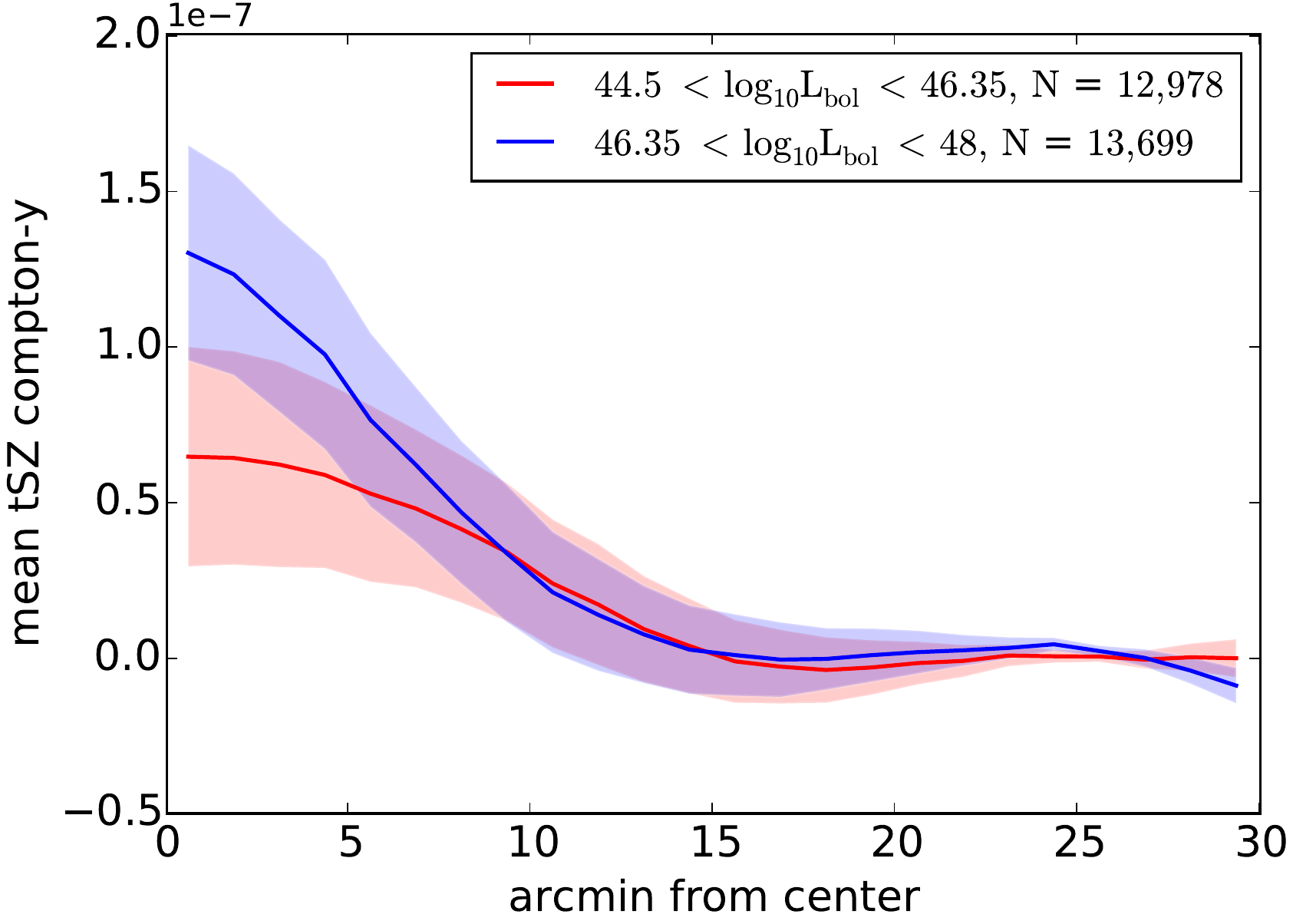} 
\includegraphics[width=0.47\textwidth]{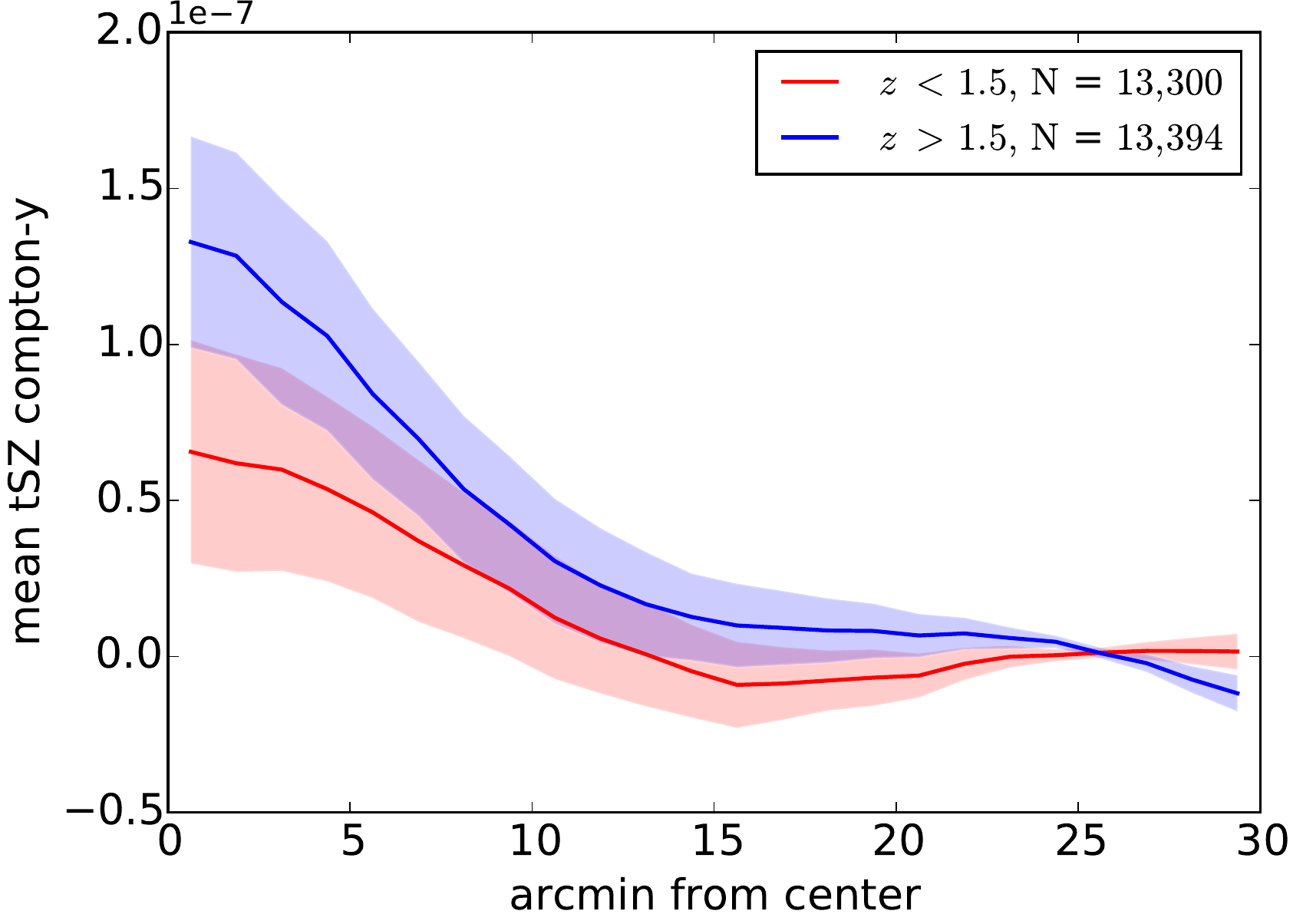} 
\caption{
Radial tSZ Compton-y profiles of mean stacked maps of quasars (all dust-corrected), separating the stacks 
into a low and high black hole mass sample (top panel), a low and high bolometric luminosity sample 
(middle panel), and a low and high redshift sample (bottom panel). A clear trend of higher tSZ Compton-y with 
increasing black hole mass and bolometric luminosities is evident. The quasar tSZ signal is even somewhat 
stronger in the high redshift sample. In our feedback picture, this trend occurs because more massive black 
holes exist in our sample at high redshifts, and it also indicates that the signal cannot be dominated by virialized gas of 
cluster-mass halos. 
}
\label{fig:profiles}
\end{figure}

\subsection{Quasar tSZ Trends with $M_{\rm{BH}}$, $L_{\rm{bol}}$, and $z$}
	In Section 4.1, we showed that the effects of quasar feedback likely dominate our observed quasar 
tSZ signal at both low and high redshifts. As a test of this quasar feedback interpretation, we investigate the 
dependence of our detected quasar tSZ signal on physical properties of the quasars, including $M_{\rm{BH}}$, 
$L_{\rm{bol}}$, and $z$. Because the cooling time of the shocked wind is long, the feedback energy 
$E_\mathrm{fb}$ we measure is likely a total thermal feedback energy released into the surrounding gas over 
the quasar lifetime, and so a positive scaling of the tSZ signal with $M_{\rm{BH}}$ and $L_{\rm{bol}}$ is 
generically expected \citep{sc04, sc08, ch08}. We divide our quasar sample in two bins of 
$M_{\rm{BH}}$ and $L_{\rm{bol}}$, and compare their tSZ Compton-y radial profiles in separate angular-scale 
stacked maps of each subsample. The spectroscopically-derived estimates of $M_{\rm{BH}}$ and 
$L_{\rm{bol}}$ we use are provided from \citet{sh11}.

	We divide our sample into two bins in $M_{\rm{BH}}$ ($7<\mathrm{log}_{10}M_{\mathrm{BH}}<9$ $M_{\odot}$, 
and $9<\mathrm{log}_{10}M_{\mathrm{BH}}<10.5$ $M_{\odot}$), as well as in $L_{\rm{bol}}$
($44.5<\mathrm{log}_{10}L_{\mathrm{bol}}<46.3$ erg s$^{-1}$, 
and $46.3<\mathrm{log}_{10}L_{\mathrm{bol}}<48$ erg s$^{-1}$). These bins are split at approximately
the median $M_{\rm{BH}}$ and $L_{\rm{bol}}$, with median $\mathrm{log}_{10}M_{\mathrm{BH}} = \{8.7, 9.3\}$
and median $\mathrm{log}_{10}L_{\mathrm{bol}}$ = \{45.9, 46.7\} in the bins, respectively; this results in similar 
signal-to-noise for each of their separate stacked maps (all dust-corrected). Figure 7 shows the Compton-y radial 
profiles of quasars with different $M_{\rm{BH}}$ (top panel) and $L_{\rm{bol}}$ (middle panel); a clear trend of increasing 
Compton-y signal with both $M_{\rm{BH}}$ and $L_{\rm{bol}}$ is evident, as expected if the observed quasar 
tSZ signal is dominated by quasar feedback. This feedback interpretation is also supported by the positive 
scaling with $L_{\rm{bol}}$, although time-variability of the quasar continuum emission for each quasar will 
increase scatter in this trend. 

	Figure 7 (bottom panel) also shows tSZ Compton-y radial profiles of our quasar sample binned in redshift 
($z<1.5$, and $z>1.5$), with median redshift $z = \{0.96, 1.96\}$ in the redshift bins, respectively. Since our
quasar sample is approximately optical flux-limited, Malmquist bias causes the higher-redshift quasar subsample to be systematically higher in $M_{\rm{BH}}$ and $L_{\rm{bol}}$. Due to the nearly redshift-independent nature of the 
tSZ effect and its trends we detect with $M_{\rm{BH}}$ and $L_{\rm{bol}}$ in quasars, the quasar tSZ signal
actually increases with redshift in our sample. The correlations between redshift, $M_{\rm{BH}}$, and 
$L_{\rm{bol}}$ can be isolated by binning our quasars in these three parameters simultaneously, but 
the resulting small subsamples causes the signal-to-noise ratio in their stacks to be too low for comparison.
We note that since the tSZ Compton-y profiles for different quasar subsamples are composed of quasars at 
different angular diameter distances and contain contributions from virialized gas in halos of different masses, 
the small differences in their angular extents are not statistically significant.

\subsection{Implications for Quasar Feedback}

	The accretion feedback efficiency of $\eta\epsilon \simeq 0.05$ (or approximately 5\% of the black hole mass) 
we measure is rather discrepant with those needed to reproduce the $M-\sigma$ relation in cosmological galaxy 
formation simulations \citep[e.g.][]{di05, si14}, which generally find that efficiencies of $\eta\epsilon \sim 0.005-0.015$ 
are required. We speculate that such large feedback efficiencies could stem from (1) the incomplete treatment of the 
multi-phase ISM in these simulations and/or (2) additional feedback from relativistic quasar jets. 

	Regarding (1), since cosmological simulations with quasar feedback cannot resolve quasar winds, 
the sub-grid models of quasar feedback utilized implicitly assume a homogenous ISM with which the quasar 
wind interacts. However, simulations of the effects of quasar winds in more realistic multi-phase ISM 
environments have shown that the hot wind-shocked gas will leak through porous low-density regions to galactic 
scales, allowing the higher-density regions to accrete and produce the observed $M_\mathrm{BH}$-$\sigma$ 
relation \citep{wa13, bo14, zu14}. These effects are particularly important in a realistic cosmological context, in 
which filamentary structures of dense infalling gas can provide a significant portion of the accreted mass onto 
the SMBH \citep{co14}. Since the quasar feedback efficiencies used in simulations are tuned to result in the correct 
observed properties of the simulated galaxies, the assumption of a homogeneous ISM can naturally lead to the 
use of small quasar feedback efficiency parameters, and correspondingly underpredicted feedback energetics 
in comparison to our observations. 

	Regarding (2), relativistic jets in high-redshift AGN may be ubiquitous \citep{fa14} and can contribute 
significantly to the feedback energetics. Most of the power in jets -- which 
derives from the angular momentum of the BH --  is eventually converted into thermal and kinetic energy, with a 
maximum power of $\sim a^2 \dot M c^2$ where $a$ is the BH spin parameter and $\dot M$ the mass accretion 
rate \citep{bl77, tc12}. Indeed, AGN jets have been observed in clusters to drive super-bubbles, some with 
estimated energetics that reach $>$$10^{62}$erg \citep{erlund06, fa14}. While it is thought that jets do not 
normally exist during the radiatively efficient quasar state \citep[although see][]{shaw12, ru14}, they are more 
commonly associated with less efficient accretion states that may have occurred at earlier times.\footnote{Since 
we select quasars that are optically bright and radio quiet, it is likely that our systems do not currently have jets.}  
In order for jets to generate the feedback energetics we report, $\sim$10$\%$ of the BH mass would have to 
accreted during radiatively inefficient states and the BHs would need sizable spin parameters during this 
previous accretion. Jets would circumvent the feedback constraints that derive from reproducing the
$M_{\rm BH}-\sigma$ relation in simulations.

	The tentative quasar feedback detection through the tSZ effect in \emph{WMAP} by \citet{ch10} also found quasar tSZ
signals significantly larger (by more than an order of magnitude) than expectations from cosmological simulations, 
although no attempt to differentiate between feedback effects and virialized halo gas was made in that study. 
Our detailed consideration of the tSZ signal from virialized halo gas in quasars shows that it is likely to be 
subdominant to the additional effects of feedback in our observed tSZ signal. Assuming a fiducial radiative 
efficiency of $\eta = 0.1$, the large feedback efficiencies of $\epsilon \simeq 0.5$ of the mean quasar bolometric 
luminosity implied by our results are also consistent with the 0.3-1.0$L_\mathrm{bol}$ commonly observed in the 
kinetic luminosity of near-relativistic quasar winds as probed through X-ray absorption lines 
\citep[e.g.][]{char09, re09, sa09, char14}. \citet{char14} argued that these large feedback efficiencies are 
evidence for magnetically-driven quasar winds, as opposed to radiative line-driving. Our estimated energetics are 
however larger than some other estimates of quasar feedback efficiencies, such as from FeLoBAL quasars \citep{faucher12b}. 

	The large thermal energy we detect in quasar wind bubbles is unlikely to be converted substantially into 
other forms and hence may show up in other tSZ measurements. \citet{sc08} considered the relic feedback
tSZ signal in lower-redshift post-quasar-phase galaxies, showing that it is similar in amplitude to 
actively-accreting quasars, and may be detectable by stacking Compton-y profiles of low-redshift 
galaxy populations. Differentiating this relic feedback tSZ signal from hot virialized gas may be 
more difficult, since these lower-redshift galaxies sit in more massive halos in which hot virialized gas 
contributes more tSZ signal than the $z\simeq1.5$ quasar host halos in our sample. However, semi-analytic 
models place low-redshift $10^{9}M_\odot$ black holes in $10^{13.5}M_\odot$ halos \citep{wyithe02}, for 
which we showed the energetics of virialized halo gas is smaller than the feedback energetics, and thus such
detections of past quasar activity in galaxies remains a possibility. Furthermore, at $z\simeq0.1$ (similar to the 
galaxy populations studied in PC13a), the quasar blastwave radius of a typical massive galaxy $\sim$10 Gyr 
after its quasar phase is $\sim$5 $E_{62}^{1/5}~$Mpc, easily resolvable with {\it Planck}. The Compton-y 
profile template of extent $\sim$$r_{\rm vir}$ used in the matched filter approach of PC13a would not capture 
most of the blastwave Compton-y signal in post-quasar-phase galaxies, and this relic tSZ feedback signature 
may have been missed in these previous investigations.

\section{Conclusion}
	The effects of quasar feedback through winds have long been suspected to shape many of the 
observed properties of massive galaxies. We have searched for the effects of quasar feedback on ionized 
gas surrounding quasars through the tSZ effect. Using a \emph{Planck} tSZ Compton-y map with explicit 
corrections for dust contamination, we stacked the Compton-y map in regions centered on a large sample of 
SDSS spectroscopic quasars and detected a tSZ signal at $>$5$\sigma$ significance. This stacking
technique is validated through separate stacks centered on SDSS galaxies of different halo masses, from
which we measured tSZ signals in agreement with \citet{planck13a}. We showed that this tSZ signal cannot 
owe to radio emission or standard parameterizations for thermal dust emission, and that it is likely to be
dominated by feedback effects rather than from virialized halo gas. We also find that the signal scales with 
$M_\mathrm{BH}$ and $L_\mathrm{bol}$ as anticipated for a quasar feedback origin.  

	We found the mean angularly integrated Compton-y of our sample of quasars to be 
$3.5\,(\pm0.7)\times10^{-6}$ Mpc$^2$, and estimated the feedback efficiency to be 5\%\,$(\pm1\%$ statistical 
uncertainty) of the mass of the black hole (although with additional systematic uncertainties stemming from
single-epoch black hole mass estimates. Our measured quasar feedback 
energetics are larger than those used in simulations that have been calibrated to reproduce observations 
of the $M_{\rm BH}-\sigma$ relation. If the large quasar feedback efficiency we infer is confirmed, 
we speculate that this could be due to (1) the simulations not resolving the multiphase ISM, which leads 
to under-predicted feedback efficiencies to reproduce the $M_{\rm BH}-\sigma$ relation, and/or (2) 
the relic of efficient radio-mode feedback in our stack, which taps into the spin energy of the black hole.

	Our investigation can be extended using the significantly larger samples of quasars and
galaxies available in current spectroscopic surveys such as the Baryon Oscillation Spectroscopic Survey 
\citep[BOSS,][]{da13} in SDSS-III, as well as future experiments such as extended BOSS (eBOSS) in SDSS-IV 
and the Dark Energy Spectroscopic Instrument \citep[DESI,][]{le13}. In particular, improved constraints on
the quasar halo mass distribution from HOD modeling using larger samples of quasars and galaxies will lead to 
better accounting of the virialized halo gas contribution to the observed quasar tSZ signal at low-redshifts, and 
thus stronger constraints on the inferred feedback parameters. Studies of correlations between the energetics 
of quasar feedback as probed by the tSZ effect and quasar multi-wavelength properties (e.g. dust obscuration, 
broad absorption lines, host galaxy star-formation) can shed light on the driving mechanism of the wind and its 
effects on the host galaxy, similar to the trends with $M_\mathrm{BH}$ and $L_\mathrm{bol}$ that we detect.  
Stacking the tSZ from different lower-redshift galaxies populations may also indicate which galaxies have had a 
previous quasar phase. Furthermore, resolved studies of the quasar feedback tSZ signal using current higher 
angular resolution CMB experiments, such as the Atacama Cosmology Telescope \citep{hi10}, the 
South Pole Telescope \citep{ca11}, the Cerro Chajnantor Atacama Telescope \citep{wo12}, and Atacama Large 
Millimeter Array \citep{wo09}, could constrain the spatial extent of feedback effects and may also be able to better 
excise troublesome Galactic and intrinsic dust emissions. If confirmed, our detection of strong quasar feedback 
will have substantive implications for our understanding of the role of quasar feedback in the evolution of 
galaxies and of the intergalactic medium.  

\acknowledgments
	We thank Colin Hill for insightful comments on a draft of this paper, the referee for useful suggestions, Yue 
Shen,  Yusra AlSayyad, James R. A. Davenport, Michael Eracleous, 
Claude-Andr\'e Faucher-Gigu\`ere, and Alexander Tcheckhovskoy for helpful discussions, and Miguel Morales for 
inspiration. JJR gratefully acknowledges recent support provided by NASA through Chandra Award Numbers 
AR9-0015X, AR0-11014X, and AR2-13007X, issued by the Chandra X-ray Observatory Center, which is operated 
by the Smithsonian Astrophysical Observatory for and on behalf of NASA under contract NAS8-03060. 

Based on observations obtained with \emph{Planck} (http://www.esa.int/Planck), an ESA science mission with instruments and contributions directly funded by ESA Member States, NASA, and Canada.

Funding for SDSS-III has been provided by the Alfred P. Sloan Foundation, the Participating Institutions, the National Science Foundation, and the U.S. Department of Energy Office of Science. The SDSS-III web site is http://www.sdss3.org/.

SDSS-III is managed by the Astrophysical Research Consortium for the Participating Institutions of the SDSS-III Collaboration including the University of Arizona, the Brazilian Participation Group, Brookhaven National Laboratory, Carnegie Mellon University, University of Florida, the French Participation Group, the German Participation Group, Harvard University, the Instituto de Astrofisica de Canarias, the Michigan State/Notre Dame/JINA Participation Group, Johns Hopkins University, Lawrence Berkeley National Laboratory, Max Planck Institute for Astrophysics, Max Planck Institute for Extraterrestrial Physics, New Mexico State University, New York University, Ohio State University, Pennsylvania State University, University of Portsmouth, Princeton University, the Spanish Participation Group, University of Tokyo, University of Utah, Vanderbilt University, University of Virginia, University of Washington, and Yale University.

Some of the results in this paper have been derived using the Python healpy implementation of the HEALPix \citep{go05} package.

\bibliography{bibref}
\bibliographystyle{apj}

\end{document}